\documentclass[onecollarge,natbib]{svjour2}
\bibpunct{[}{]}{;}{n}{}{,} % to get "[numbered]" references from natbib
\smartqed  % flush right qed marks, e.g. at end of proof
\usepackage{graphicx}
\usepackage{bbm}
\usepackage{bm}
\usepackage{amssymb}
\journalname{Few-body Systems}
\begin{document}

\title{Techniques to treat the continuum applied to electromagnetic
  transitions in $^8$Be}

\author{E. Garrido \and A.S. Jensen \and D.V. Fedorov }

\institute{E. Garrido \at Instituto de Estructura de la Materia, IEM-CSIC, Serrano 123, E-28006 Madrid, Spain \\
     \email{e.garrido@csic.es}           %  \\
     \and
     A.S. Jensen \at Department of Physics and Astronomy, Aarhus University, DK-8000 Aarhus C, Denmark \\
     \and
     D.V. Fedorov \at Department of Physics and Astronomy, Aarhus University, DK-8000 Aarhus C, Denmark \\
}

\date{Received: date / Accepted: date}

\maketitle

\begin{abstract}
Bremsstrahlung emission in collisions between charged nuclei is
equivalent to nuclear gamma decay between continuum states. The way
the continuum spectrum can be treated is not unique, and efficiency
and accuracy of cross section calculations depend on the chosen
method.  In this work we describe, relate, and compare three different methods 
in practical calculations
of inelastic cross sections, that is, by (i) treating the initial and
final states as pure continuum states on the real energy axis, (ii)
discretizing the continuum states on the real energy axis with a box
boundary condition, and (iii) complex rotation of the hamiltonian
(complex scaling method). The electric quadrupole transitions, $2^+
\rightarrow 0^+$ and $4^+ \rightarrow 2^+$, in $\alpha+\alpha$
scattering are taken as an illustration. 
\end{abstract}

\section{Introduction}

The emission of bremsstrahlung in a collision between two charged particles constitutes an important
background effect in Coulomb deexcitation processes. In a classical picture, this phenomenon is
understood as the energy radiated due to the deceleration of a charged particle when deflected
by another charged particle. The radiated energy is just the kinetic energy lost in the deceleration
process. In a quantum mechanical picture, the process can be seen as the $\gamma$-emission due to
the decay of a two-body system from some two-body continuum state into another continuum state of lower 
energy. A detailed derivation of the cross section for this kind of processes can be found in \cite{ald56}.

The fact that these transitions involve continuum structures, both in
the initial and final states, introduce technical as well as
conceptual difficulties in the calculations.  First, the treatment of
the continuum spectrum itself, which can be handled in different ways,
but in numerical studies almost necessarily by some kind of
discretization.  Second, the unavoidable matrix elements between the
continuum states involved in the calculation are diverging and thus
not well defined without some regularization prescription.  These two
problems, how to treat the continuum and how to obtain converged
results, are possible sources of uncertainty that deserves detailed
investigations.

Different methods were previously used to compute the bremsstrahlung
cross sections for various combinations of nuclei.  In \cite{lan86}
$\alpha-\alpha$ collisions were investigated in connection with the
resonant electric quadrupole capture into the unbound ground state of
$^8$Be.  The bremsstrahlung cross section was obtained without any
discretization of the continuum spectrum.  The two-body states are
then pure continuum structures, which are orthogonal to each other in
a continuum sense, through a Dirac delta.

Another possibility is for instance employed in \cite{ber99}, where the
bremsstrahlung radiation during $\alpha$-decay was computed in a time
dependent picture for heavy nuclei.  In this case the continuum states
were discretized and treated on the same footing as the discrete part
of the spectrum describing the bound states. The orthogonality
condition is now given by a Kronecker delta. In both these two
procedures the energy is a real number, which means that the
resonances (if any) are not isolated as particular continuum states,
and their effect on the cross section is diluted into the continuum
spectrum (discretized or not).

In \cite{myo98} the Coulomb breakup reaction of $^{11}$Be was analyzed
by use of the complex scaling method.  This method \cite{ho83} rotates
the usual coordinates into the complex plane ($x \rightarrow x
\exp(i\theta)$), and permits an easy separation between resonances and
ordinary continuum states.  The resonances, defined as poles of the
${\cal S}$-matrix, appear as discrete solutions clearly separated from
the (rotated) continuum background, provided the rotation angle is
sufficiently large.  This method provides simultaneously the complex
resonance energy with real part and imaginary part equal to minus half
the width.  The complex energy of the resonance can also be found by extending the
energy into the complex plane without any complex rotation of the
coordinates.  The complex scaling method has the enormous advantage
that the resonance wave function falls off exponentially at large
distances, exactly as ordinary bound states. This fact permits to
circumvent all the technical difficulties arising from the otherwise
exponentially divergent resonance wave function.  However, the
continuum background still contributes and must therefore necessarily
be included in calculations of the observable cross sections.

The $^8$Be nucleus is particularly interesting due to the role it most
likely plays in the triple-alpha radiative capture into $^{12}$C,
which is one of the most important reactions in stellar
nucleosyntheses.  This nucleus is rather well described as an
$\alpha$+$\alpha$ molecular cluster structure, and therefore equally
well treated as a two-body problem.  In a recent work \cite{gar13} the
full continuum method was used to investigate the $E2$-transitions in
$^8$Be. In particular, the cross section was found to be insensitive
to the $\alpha$-$\alpha$ interaction used, and it was also shown how a
precise definition of the cross section requires a choice of an energy
window for the final states in the scattering process.

The existence and energy sequence of the $0^+$, $2^+$, and $4^+$
resonances suggest the interpretation of these states as a rotational
band.  However, these states are located in the continuum with
corresponding decay widths.  In other words, continuum properties are
an integral part of understanding $^8$Be.  Rotational bands in the
continuum present conceptual problems arising from substantial decay
widths of these states. This was recently discussed in \cite{gar13a}
where the (non-observable) structure dependent electromagnetic
transition probabilities approximately were extracted from
(observable) cross sections.

Despite the apparent simplicity of $^8$Be this nucleus is not yet
fully understood, because all properties are continuum related, since
even the ground state is unbound.  Furthermore, $^8$Be is the only
spontaneously fissioning nucleus along the beta-stability line before
Uranium.  This fission process is complicated due to both dynamics and
delicate balancing of various energy terms.  On the other hand, the
simplicity itself makes $^8$Be an almost perfect as a test case, since
it still is complex enough to maintain essential key features.  These
should be understood before the properties of more complicated systems
can be fully appreciated and exploited.

The purpose of the present work is to describe and compare the three
methods mentioned in the paragraphs above.  We can then pinpoint when
a given method is preferable for computation and interpretation of a
given quantity.  We can also specifically distinguish between results
for the same quantities computed by different methods for the same
nucleus.  We shall refer to the three methods as the full continuum
method, the discretized continuum method, and the complex scaling
method. The main aspects of each of them will be given in sections
\ref{sec2}, \ref{sec3}, and \ref{sec4}, respectively. The connection
between all the three methods will be shown.  Each of the three
sections contains a subsection where the corresponding method is used
to describe the electric quadrupole transitions, $2^+ \rightarrow 0^+$
and $4^+ \rightarrow 2^+$, in $^8$Be, which are taken as an
illustration.  We close the paper with the summary and the
conclusions.

\section{Full continuum method}
\label{sec2}

A detailed derivation of the bremsstrahlung cross section for the collision between two charged
particles can be found in Ref.~\cite{ald56}. To be
precise, the final result for the differential cross section is given in Eq.(II.3.2) of that reference.
A summary of the most relevant expressions can also be found in Ref.~\cite{tan85}, where the expressions
for the angular integrated cross sections are also given.
We first give the expressions for spin-zero bosons which afterwards is
applied to collisions between two alpha-particles

\subsection{General formulation for spin-zero bosons}

The dominating electromagnetic transitions between states of two
spin-zero bosons are necessarily of quadrupole character.  This
applies in particular to the two-alpha system, where Eq.(11) in
\cite{tan85} takes the form:
\begin{equation}
\left.
\frac{d\sigma}{dE_\gamma} \right|_{\ell \rightarrow \ell^\prime}(E)=
\frac{4\pi^2 e^2}{15 k^2}\left(\frac{E_\gamma}{\hbar c} \right)^5 
(2\ell+1)  \left|
\langle \ell 0; 2 0| \ell^\prime 0\rangle
\int_0^\infty u_\ell(E,r) r^2 u_{\ell^\prime}(E^\prime,r) dr
\right|^2,
\label{cross1}
\end{equation}
where $E$ and $E^\prime$ are the initial and final state energies in
the two-body center of mass frame, $E_\gamma=E-E^\prime$ is the energy
of the emitted photon, $\ell$ and $\ell^\prime$ are the relative
angular momenta between the two particles in the initial and final
state, $e$ is the unit charge, and $k^2=2 \mu E/\hbar^2$, where $\mu$
is the reduced mass of the two-body system.  In Eq.(\ref{cross1}) we
used a charge of $2e$ for each alpha particle amounting to the factor
of $4e^2$.

An important point refers to the radial two-body wave functions $u_\ell$ and $u_{\ell^\prime}$. They are
the solutions of the radial two-body Schr\"{o}dinger equation:
\begin{equation}
\left[
-\frac{\hbar^2}{2\mu}\frac{d^2}{dr^2}+\frac{\hbar^2}{2\mu}\frac{\ell(\ell+1)}{r^2}+V(r)-E
\right]u_\ell(E,r)=0
\label{schr}
\end{equation}
where $V(r)$ is the two-body interaction. These wave functions behave asymptotically
as:
\begin{equation}
u_\ell(E,r) \stackrel{r \rightarrow \infty}{\longrightarrow} 
C \left[ \cos\delta_\ell F_\ell(kr) + \sin\delta_\ell G_\ell(kr)\right],
\label{asymp}
\end{equation}
where $F_\ell$ and $G_\ell$ are the regular and irregular Coulomb functions, $\delta_\ell$ is the nuclear
phase shift, and the asymptotic constant $C$ is determined from the energy normalization condition, which 
requires that:
\begin{equation}
\int_0^\infty u_\ell(E,r) u_\ell(E^\prime,r) dr=\delta(E-E^\prime).
\label{ener}
\end{equation}
This normalization condition implies that the asymptotic constant $C$ has to be: 
\begin{equation}
C=\sqrt{\frac{2 \mu}{\pi \hbar^2 k}}.
\label{constant}
\end{equation}
A derivation of the value of the asymptotic constant can be found in
appendix~\ref{app1}. Note that from Eqs.(\ref{asymp}) and
(\ref{constant}) we have that the units of the radial continuum wave
functions $u$ are one over square root of Energy times Length, which is also
consistent with the normalization condition (\ref{ener}), and which
leads to the correct units of length squared divided by energy for the
energy differential cross section in Eq.(\ref{cross1}).

The total bremsstrahlung cross section, as a function of the incident energy $E$,  is obtained
after integration over the energy of the emitted photon:
\begin{equation}
\sigma(E)=\int \left. 
\frac{d\sigma}{dE_\gamma}\right|_{\ell \rightarrow \ell^\prime} \hspace*{-5mm}(E) \;dE_\gamma.
\label{intcs}
\end{equation}
As stated in Ref.\cite{gar13}, the computed cross sections should be obtained in
close analogy to the experimental setup, where only a finite range of
final relative energies is measured. This means that the integral in
Eq.(\ref{intcs}) has to be performed only over this precise energy
range. We shall often refer to this range as the final energy window.

A delicate point in the calculation of the cross section refers to the
procedure employed to obtain the radial integral in
Eq.(\ref{cross1}). Due to the fact that the continuum wave functions
do not converge towards zero at infinity (see Eq.(\ref{asymp})), the
integrand in Eq.(\ref{cross1}) actually diverge with distance by
oscillating with larger and larger amplitude. Some regularization
prescription is required to extract the physically meaningful content.
This is possible since the matrix elements physically must be well
defined, and mathematically as well corresponding to cancellation of
the large-distance contributions. A convenient numerical procedure
must then be found and applied.

In this work we shall use the Zel'dovich regularization \cite{zel60},
which introduces the regularization factor $e^{-\eta^2 r^2}$ in the
diverging radial integrand.  Then the desired correct result is
obtained in the limit of zero value for the Zel'dovich parameter
$\eta$.  This removes the unwanted large-distance large-amplitude
oscillations and the uniquely defined limiting result is obtained for
sufficiently small, but finite, values of $\eta$.  The smaller the
value of $\eta$ the slower the fall off of the radial integrand, and
therefore the larger the upper limit required in the radial integral
in Eq.(\ref{cross1}).  Numerically, this obviously becomes more and
more difficult since the large-amplitude oscillations cancel each
other and consequently must be very accurately computed.  The optimum
value of $\eta$ is therefore as large as possible yet sufficiently
small to have reached the limit. A different choice is of course possible 
for the regularization factor. For instance one could use a higher power in 
the exponent, which would produce a faster attenuation of the tail of the 
integrand. However, the use a higher power would reduce the range interval 
of $\eta$-values in which the integral stabilizes at the correct result. The choice 
made in this work is a sort of compromise between how fast the unwanted tail
in the integrand is killed and the difficulty of finding a range of $\eta$-values 
in which the integral reaches the $\eta=0$ value with sufficient accuracy.

\subsection{Full continuum wave functions: $E2$-capture in $\alpha+\alpha$ collisions.}

The expressions given in the subsection above provide a summary of the
most pertinent formulae for computations of bremsstrahlung cross
sections.  They are from Refs.~\cite{lan86,lan86b}, where the $2^+
\rightarrow 0^+$ and $4^+ \rightarrow 2^+$ transitions in $^8$Be were
computed.  In \cite{lan86} the Buck $\alpha-\alpha$ potential given in
\cite{buc77} was used, while in \cite{lan86b} the results obtained
with the Buck potential and the Ali-Bodmer potential \cite{ali66} were
compared. The main difference between these potentials is in the
treatment of the Pauli principle.  The Buck potential generates a
nodal structure in the two-body wave functions in accordance with
microscopic theories \cite{buc77}. The immediate consequence is the
appearance of two bound $0^+$-states and one bound $2^+$-state in
$^8$Be (with energies $-72.79$ MeV, $-25.88$ MeV, and $-22.28$ MeV,
respectively). These spurious states correspond to Pauli forbidden
states.  On the other hand the Ali-Bodmer potential contains a
short-distance repulsion that prevents the appearance of such
forbidden states. Both potentials reproduce equally well the $\ell=0$,
$\ell=2$, and $\ell=4$ phase shifts (up to $E\sim 20$ MeV), and
therefore the corresponding two-body wave functions have the same
asymptotic behavior.

Although in \cite{lan86b} the cross section was found to depend quite
a lot on the potential used, in subsequent calculations \cite{kro87}
this dependence was reduced significantly. In fact, in our recent work
\cite{gar13} we have shown that the dependence is actually very small,
and the cross sections obtained with the Ali-Bodmer and Buck
potentials (and even with the phase equivalent version of the Buck
potential) can be hardly distinguished. For this reason in this work
any discussion about the dependence on the potential will be omitted,
and we will show the results obtained with the Buck potential only. In
particular, with this potential the $0^+$, $2^+$, and $4^+$ resonances
are found at 0.091 MeV, 2.88 MeV, and 11.78 MeV, respectively, and
their corresponding widths are 3.6 eV, 1.24 MeV, and 3.57 MeV. These
values agree very well with the experimental values given in
\cite{til04}.

\begin{figure}
\centering
\includegraphics[width=10cm]{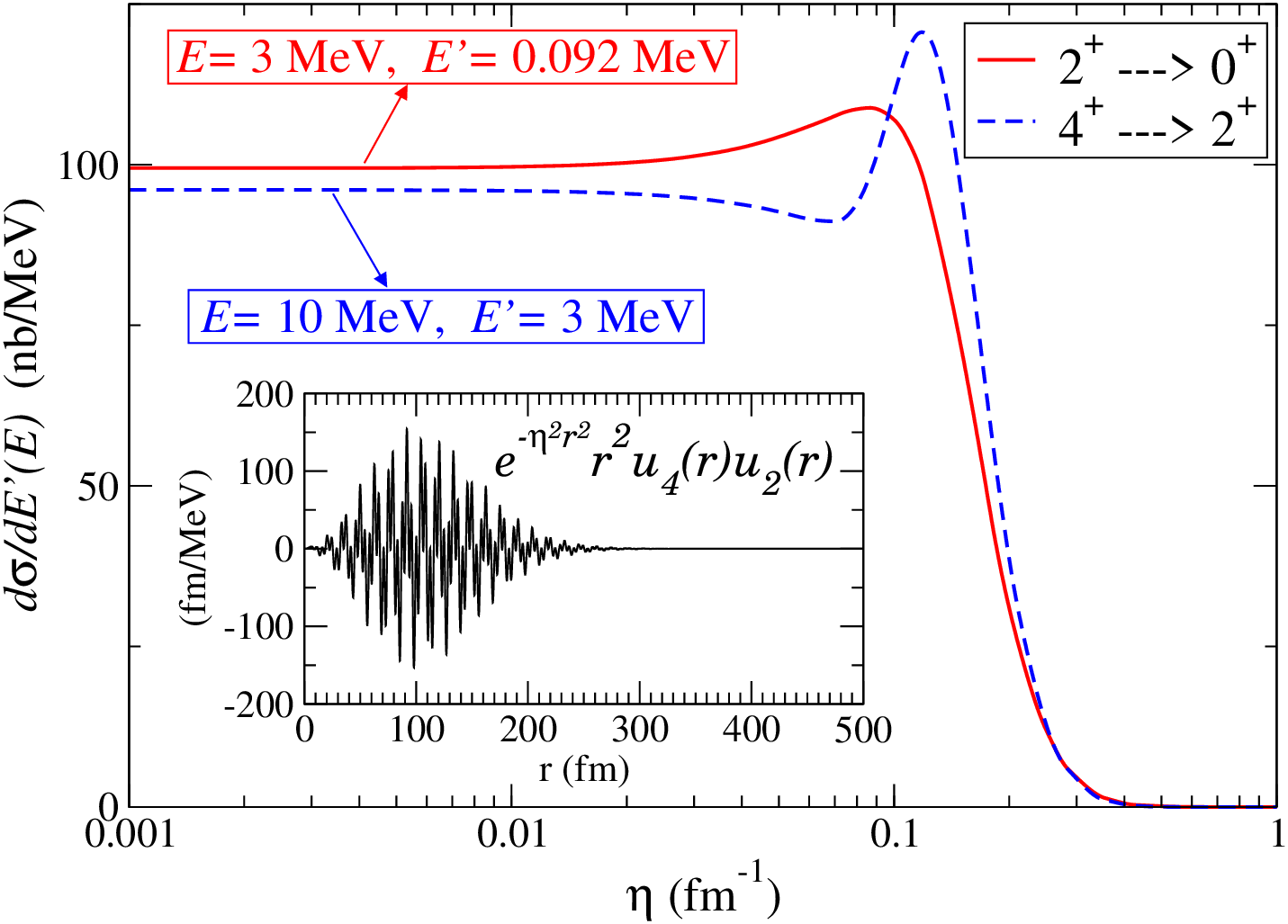}
\caption{(Color online) Outer part: Differential cross section
  (Eq.(\ref{cross1})) for given initial and final energies ($E$ and
  $E^\prime$) for the $2^+ \rightarrow 0^+$ (solid curve) and $4^+
  \rightarrow 2^+$ (dashed curve) transitions in $^8$Be as a function
  of the Zel'dovich parameter $\eta$ used to regularize the integrand
  in the radial integral. The values of the initial and final energies
  are $E=3.0$ MeV and $E^\prime=0.092$ MeV for the $2^+ \rightarrow
  0^+$ transition, and $E=10.0$ MeV and $E^\prime=3.0$ MeV for the
  $4^+ \rightarrow 2^+$ transition, respectively.  Inner part:
  Integrand in the radial integral in Eq.(\ref{cross1}) corresponding
  to the differential cross section for the $4^+ \rightarrow 2^+$
  transition shown in the outer part for $\eta=0.01$ fm$^{-1}$.}
\label{fig1}       
\end{figure}

Let us first investigate the dependence of the results on the
Zel'dovich parameter $\eta$. As mentioned above, this parameter enters
in the regularization factor $e^{-\eta^2 r^2}$ used to extract the
physics from the wildly oscillating integrand in Eq.(\ref{cross1}).
In Fig.~\ref{fig1} we show the differential cross section
(\ref{cross1}) for specific values of the initial and final energies
$E$ and $E^\prime$ as a function of $\eta$. The solid and dashed
curves in the figure correspond to the $2^+ \rightarrow 0^+$
transition ($\ell=2, \ell^\prime=0$ in Eq.(\ref{cross1})) and the $4^+
\rightarrow 2^+$ transition ($\ell=4, \ell^\prime=2$ in
Eq.(\ref{cross1})), respectively. The chosen values for $E$ and
$E^\prime$ are 3.0 MeV and 0.092 MeV for the $2^+ \rightarrow 0^+$
transition, and 10.0 MeV and 3.0 MeV for the $4^+ \rightarrow 2^+$
transition. As seen in the figure, the cross sections are very stable
for sufficiently small values of $\eta$. This proves that the $\eta
\rightarrow 0$ limit is properly reached. Values of $\eta$ smaller
than about 0.02~fm$^{-1}$ then provide the converged value of the
cross section in the $\eta \rightarrow 0$ limit.

Furthermore, keeping $\eta \sim 0.01$ fm$^{-1}$ the upper radial limit
required in the integral in Eq.(\ref{cross1}) stays within reasonable
values, such that the integral is not difficult to handle. As an
example, we show in the inner part of Fig.\ref{fig1} the integrand
corresponding to the $4^+ \rightarrow 2^+$ transition shown in the
outer part for $\eta=0.01$ fm$^{-1}$. As we can see, an upper limit of
about 400$\sim$500 fm is enough. In case of using for instance
$\eta=0.001$ fm$^{-1}$ the upper limit moves up til about 5000 fm, and
integration of such a highly oscillating function up to that distance
becomes a much more delicate task.  Of course, the larger the value of
$\eta$ the more the integrand is killed, and eventually, for
sufficiently large values of $\eta$ the computed differential cross
section approaches zero.  All the results shown later on in this work
will be obtained with $\eta=0.01$ fm$^{-1}$.  We emphasize that the
strongly oscillating integrand is a result of the two regularly
oscillating continuum wave functions $(u_{\ell},u_{\ell'})$.  The
strong cancellation is a consequence, and the resulting well-defined
value of the integral is orders of magnitude smaller than
corresponding to the amplitude of the oscillations at the moderate
distances, see the inset in Fig.\ref{fig1}.

We have then computed the total bremsstrahlung cross section for the
$2^+ \rightarrow 0^+$ and $4^+ \rightarrow 2^+$ transitions according
to Eqs.(\ref{cross1}) and (\ref{intcs}). The $0^+$, $2^+$, and $4^+$
continuum wave functions are computed numerically by solving
Eq.(\ref{schr}) for an arbitrarily small grid of energies $E$ and
$E^\prime$ ($E_\gamma=E-E^\prime > 0$). The computed wave functions
are scaled such that the asymptotic behavior is given by
Eqs.(\ref{asymp}) and (\ref{constant}).  Another important ingredient
in the calculation is the final energy window for the integral in
Eq.(\ref{intcs}). For the $2^+ \rightarrow 0^+$ transition it was
shown in Ref.\cite{gar13} that, due to the very small width of the
$0^+$ resonance in $^8$Be, a final energy window for the $0^+$ states
of 0.5 keV around the $0^+$ resonance energy is enough to reach
convergence for the cross section.  This width for the window is far
smaller than the best experimental resolution of about 10 keV.  For
the $4^+ \rightarrow 2^+$ transition the cross section is much more
sensitive to the size of the final energy window chosen in
Ref.\cite{gar13}.  In the present work we shall use the same window as
in \cite{lan86b,kro87,lan86c}, namely, $2 \mbox{ MeV} < E' < 4 \mbox{
  MeV}$, which roughly corresponds to the $2^+$-resonance energy $\pm
1$ MeV and also is comparable to its widths of $1.24$~MeV.

\begin{figure}
\centering
\includegraphics[width=10cm]{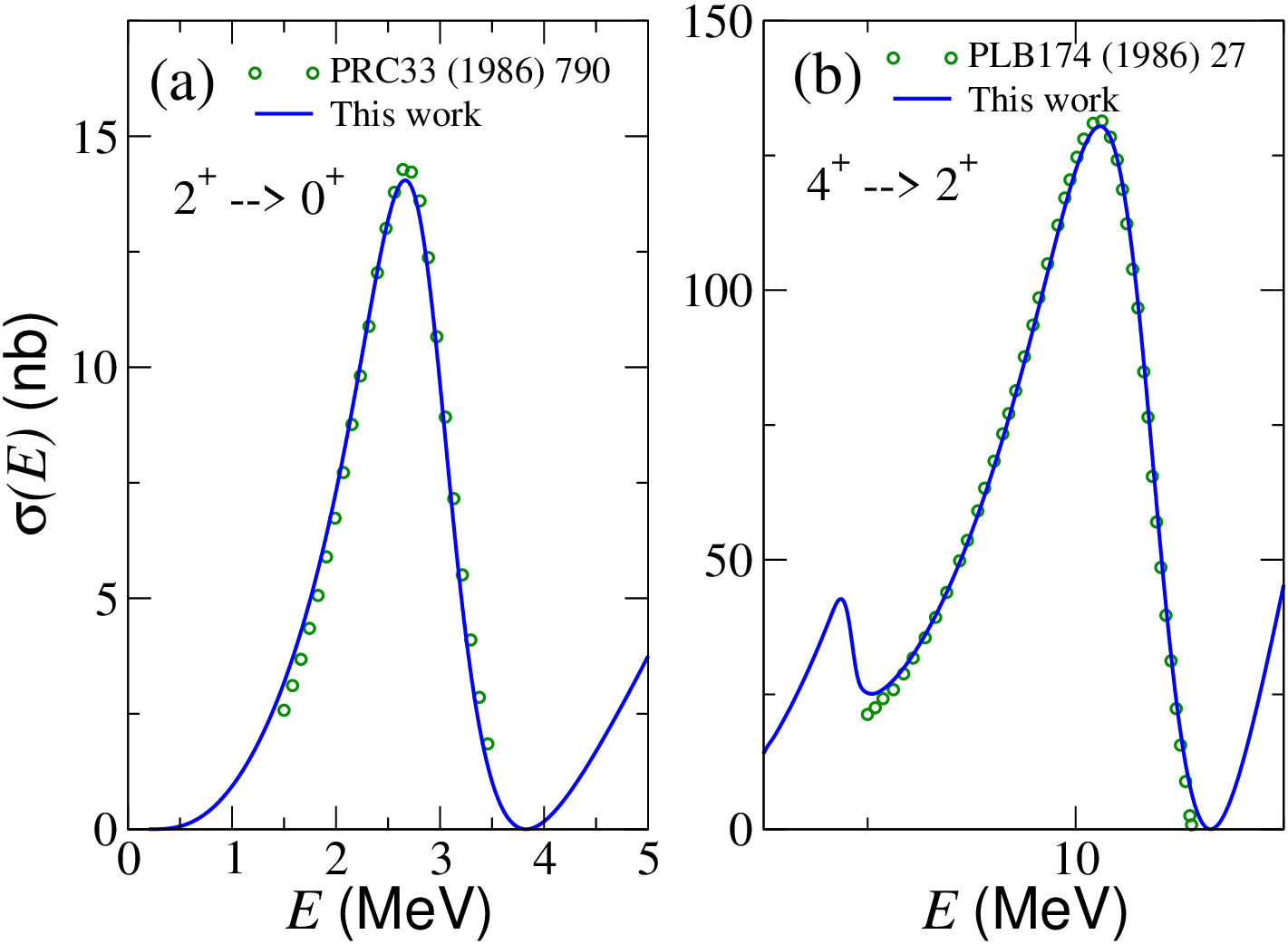}
\caption{(Color online) Integrated bremsstrahlung cross section
  (Eq.(\ref{intcs})) for the (a) $2^+ \rightarrow 0^+$ and (b) $4^+
  \rightarrow 2^+$ transitions in $^8$Be as a function of the incident
  energy $E$.  The integral in  Eq.(\ref{intcs}) has been performed over the
  final energy windows specified in the text.  The open circles in
  parts (a) and (b) correspond to the cross section obtained with the
  Buck potential in Refs.~\cite{lan86} and \cite{lan86b},
  respectively. }
\label{fig2}       
\end{figure}

In Fig.~\ref{fig2} the solid lines show the computed total
bremsstrahlung cross section integrated over the corresponding energy
windows (Eq.(\ref{intcs})) for the $2^+ \rightarrow 0^+$
(Fig.\ref{fig2}a) and $4^+ \rightarrow 2^+$ (Fig.\ref{fig2}b)
transitions in $^8$Be. In both figures the open circles are the
results obtained with the Buck potential used in Refs.\cite{lan86} and
\cite{lan86b}.  The procedure described in this section is very
similar to the one used in these two references, and therefore it is
not surprising to find the good agreement between our results and the ones in
\cite{lan86,lan86b}. However, it is important to keep in mind that
when using a different $\alpha-\alpha$ potential (for instance the
Ali-Bodmer potential), we obtain a very similar cross section, while
in \cite{lan86b,kro87} a big discrepancy was found for the $4^+
\rightarrow 2^+$ transitions (see \cite{gar13} for details). The peak
at low energies observed in Fig.\ref{fig2}b is due to the known
$1/E_\gamma$ dependence of the bremsstrahlung cross section at small
photon energies \cite{gre01}. This is the so called infrared
catastrophe.  However, as explained in \cite{gre01}, this divergence
is not physical. A transition with $E_\gamma=0$ is nothing but an
elastic process. A relativistic treatment of the elastic reaction up
to the same order will produce a similar $1/E_\gamma$ divergence in
the cross section but with opposite sign that precisely cancels the
one obtained in the calculation of the bremsstrahlung cross section.
As shown in \cite{gar13}, removal of the soft-photon contributions
removes as well the nonphysical peak in the cross section.

\section{Discretized continuum method}
\label{sec3}

Instead of using continuum wave functions with the energy
normalization in Eq.(\ref{ener}), a quite common procedure is to
discretize the continuum spectrum into an orthogonal set of basis
functions, and treat them similarly to bound states. One option,
rather often used, is to average the continuum states over narrow
range of energies. For each of these energy bins a discrete bin wave
function is constructed \cite{ber96,nun99}. These wave functions are
automatically normalized to 1 provided that the radius of the bin wave
function is large enough. In this work we shall employ the much
simpler procedure of imposing a box boundary condition. In this way
the spectrum is automatically discretized, and each state is then
normalized to 1.  We first give the general expressions in some
details, and then apply to collisions of two alpha-particles.

\subsection{General formulation}

The connection between the continuum wave functions used in the previous section and the discrete
ones can be seen rather easily. To do it, let us assume that we have discretized the continuum spectrum 
in a box of size $L$. We then get a family
of discrete states $\left\{ u_\ell^{(i)}(E_i,r) \right\}$ satisfying that
\begin{equation}
\int_0^L u_\ell^{(i)}(E_i,r) u_\ell^{(j)}(E_j,r) dr =\delta_{ij}.
\label{orton}
\end{equation}
Each state $i$ corresponds to some discrete energy value $E_i$, and it satisfies the box boundary condition  
$u_\ell^{(i)}(E_i,L)=0$. The set of energy values $\{E_i\}$ is then not arbitrary, but dictated
by the size of the box. The number of discrete states grows linearly with $L$, and therefore, the
larger $L$ the smaller the energy separation between the discrete states.

Asymptotically each function $u_\ell^{(i)}(E_i,r)$ behaves as $\sim \sin(k_i r+\delta)$, where $\delta$
is the phase shift and 
\begin{equation}
k_i^2=2\mu E_i/\hbar^2. 
\label{mom}
\end{equation}
Imposing now that $u_\ell^{(i)}(E_i,L)=0$, we get that the discrete values of the momentum $k_i$ take 
the form:
\begin{equation}
k_i\approx i \pi/L; \hspace{1cm} (i=1,2,\cdots),
\label{discr}
\end{equation}
from which, and making use of Eq.(\ref{mom}), it is not difficult to see that the energy separation between 
two consecutive discrete states is given by:
\begin{equation}
\Delta E=E_i-E_{i-1}=(k_i^2-k_{i-1}^2) \frac{\hbar^2}{2\mu}\approx 
\frac{2\pi}{L} \frac{\hbar^2 k_i}{2 \mu}=\frac{2\pi}{L} \frac{E_i}{k_i},
\label{deltae} 
\end{equation}
where we have assumed that $i$ is large enough such that
$(i^2-(i-1)^2)\approx 2i$. From the expression above it is now evident
that the energy distance between two consecutive energies decreases
linearly with $L$.

Let us now use the relation between the Dirac and Kronecker deltas:
\begin{equation}
\delta(E_i-E_j)=\lim_{\Delta E \rightarrow 0} \frac{\delta_{ij}}{\Delta E},
\label{deltas}
\end{equation}
where $\Delta E$ is the separation between the two energies.  From this expression the normalization
conditions (\ref{orton}) and (\ref{ener}) can be easily related:
\begin{equation}
\int_0^\infty u_\ell(E_i,r) u_\ell(E_j,r) dr= 
\lim_{L\rightarrow \infty} 
\int_0^L \frac{u_\ell^{(i)}(E_i,r)}{\sqrt{\Delta E}} \frac{u_\ell^{(j)}(E_j,r)}{\sqrt{\Delta E}} dr,
\end{equation}
where the wave functions $u_\ell(E,r)$ are continuum functions in the sense of section~\ref{sec2}, and where 
we have used that, as seen in Eq.(\ref{deltae}), to impose $\Delta E \rightarrow 0$ amounts
to imposing $L\rightarrow \infty$.

Making use again of Eq.(\ref{deltae}), we can relate the pure continuum
wave functions and the discrete wave functions. This relation is given by:
\begin{equation}
u_\ell(E_i,r) = 
\!\! \lim_{\Delta E\rightarrow 0} \frac{u_\ell^{(i)}(E_i,r)}{\sqrt{\Delta E}} =
\!\! \lim_{L\rightarrow \infty} \sqrt{\frac{L}{2}} \sqrt{\frac{2\mu}{\pi\hbar^2k_i}}
                                           u_\ell^{(i)}(E_i,r).
\label{relat}
\end{equation}
From this expression one can see that while the units of the continuum
wave functions $u_\ell$ are one over square root of Energy times
Length, (consistent with (\ref{ener})), the discrete states
$u_\ell^{(i)}$ have units of one over square root of length
(consistent with (\ref{orton})).

Furthermore, taking into account the asymptotic behavior of the continuum wave functions, which 
is  given by Eqs.(\ref{asymp}) and (\ref{constant}), it is now evident that the discretized continuum states
satisfy:
\begin{equation}
u_\ell^{(i)}(E_i,r) \stackrel{r \rightarrow \infty}{\longrightarrow} 
\sqrt{\frac{2}{L}} \left[ \cos\delta_\ell F_\ell(k_ir) + \sin\delta_\ell G_\ell(k_i r)\right].
\label{asymp2}
\end{equation}

Now, in order to compute the bremsstrahlung cross section
Eq.(\ref{cross1}), the only remaining point is to translate the radial
integral in this equation into the discrete continuum spectrum
language. To do so, let us first write the square of the radial
integral in a more compact way as:
\begin{equation}
\left|
\int_0^\infty u_\ell(E,r) r^2 u_{\ell^\prime}(E^\prime,r) dr
\right|^2=
\langle u_\ell(E,r) | r^2 | u_{\ell^\prime}(E^\prime,r) \rangle
\langle u_{\ell^\prime}(E^\prime,r) | r^2 | u_\ell(E,r) \rangle.
\label{radial}  
\end{equation}

Since the discrete continuum states form a complete basis, we can now exploit the completeness relation
\begin{equation}
\mathbbm{1}=\sum_i |u_\ell^{(i)}(E_i,r)\rangle \langle u_\ell^{(i)}(E_i,r) |,
\end{equation}
which of course applies also to the final states with relative orbital angular momentum $\ell^\prime$.
If we now insert the initial state unity operator in between the $u_\ell$ functions and $r^2$, and
similarly, the final state unity operator in between the $u_{\ell'}$ functions and $r^2$,
we can then rewrite Eq.(\ref{radial}) as:
\begin{eqnarray}
\left|
\int_0^\infty u_\ell(E,r) r^2 u_{\ell^\prime}(E^\prime,r) dr
\right|^2 = & &
\sum_{i j}
\langle u_\ell(E,r) | u_\ell^{(i)}(E_i,r)\rangle 
\langle u_\ell^{(i)}(E_i,r) |r^2 | u_{\ell'}^{(j)}(E^\prime_j,r) \rangle
\langle u_{\ell'}^{(j)}(E^\prime_j,r)  |  u_{\ell^\prime}(E^\prime,r) \rangle 
\nonumber \\ & & \hspace*{-2cm} \times
\sum_{i' j'}
\langle u_{\ell^\prime}(E^\prime,r) | u_{\ell'}^{(j')}(E^\prime_{j'},r) \rangle
\langle u_{\ell'}^{(j')}(E^\prime_{j'},r) | r^2 | u_{\ell}^{(i')}(E_{i'},r) \rangle
\langle  u_{\ell}^{(i')}(E_{i'},r)  | u_\ell(E,r) \rangle,
\label{radial2} 
\end{eqnarray}
where the energies with and without primes refer, respectively, to energies in the initial and final states.

Since Eq.(\ref{relat}) can also be written as:
\begin{equation}
u_\ell^{(i)}(E_i,r)=\lim_{\Delta E \rightarrow 0} \sqrt{\Delta E} u_\ell(E_i,r), 
\end{equation}
we then have that
\begin{equation}
\langle u_\ell(E,r) | u_\ell^{(i)}(E_i,r)\rangle=
\lim_{\Delta E \rightarrow 0} \sqrt{\Delta E} \; \delta(E-E_i),
\end{equation}
and Eq.(\ref{radial2}) becomes:
\begin{eqnarray}
\left|
\int_0^\infty u_\ell(E,r) r^2 u_{\ell^\prime}(E^\prime,r) dr
\right|^2 &= &
\lim_{\Delta E \rightarrow 0}
\sum_{i j} 
\Delta E \delta(E-E_i) \delta(E'-E^\prime_j)
\langle u_\ell^{(i)}(E_i,r) |r^2 | u_{\ell'}^{(j)}(E^\prime_j,r) \rangle
\nonumber \\ &\times&
\sum_{i' j'}
\Delta E \delta(E'-E^\prime_{j'}) \delta(E-E_{i'})
\langle u_{\ell'}^{(j')}(E^\prime_{j'},r) | r^2 | u_{\ell}^{(i')}(E_{i'},r) \rangle.
\label{radial3}
\end{eqnarray}

Finally, from Eq.(\ref{deltas}) we have that 
\begin{equation}
\lim_{\Delta E \rightarrow 0} \Delta E \delta(E-E_i)\delta(E-E_{i'}) =
\lim_{\Delta E \rightarrow 0} \Delta E \delta(E-E_i)\delta(E_i-E_{i'})=\delta(E-E_i)\delta_{ii'},
\end{equation}
and two of the summations in Eq.(\ref{radial3}) can be trivially made, which leads 
to the final result:
\begin{equation}
\left|
\int_0^\infty u_\ell(E,r) r^2 u_{\ell^\prime}(E^\prime,r) dr
\right|^2=
 \sum_{ij} \delta(E-E_i) \delta(E^\prime-E^\prime_j)
\left|
\langle u_\ell^{(i)}(E_i,r) |r^2 | u_{\ell'}^{(j)}(E^\prime_j,r) \rangle
\right|^2,
\label{radial4}
\end{equation}
where
\begin{equation}
\langle u_\ell^{(i)}(E_i,r) |r^2 | u_{\ell'}^{(j)}(E^\prime_j,r) \rangle
=
\int_0^L
u_\ell^{(i)}(E_i,r) r^2  u_{\ell'}^{(j)}(E^\prime_j,r) dr.
\label{integ2}
\end{equation}

Therefore, in the discretized continuum picture, the differential bremsstrahlung cross section is given by
Eq.(\ref{cross1}), where the square of the radial integral is given by (\ref{radial4}). Thanks to the 
delta functions the integral (\ref{intcs}) can be trivially made, and we get for the integrated cross section:
\begin{equation}
\sigma(E)= \frac{4\pi^2 e^2}{15 k^2} (2\ell+1)  
\langle \ell 0; 2 0| \ell^\prime 0\rangle^2
\sum_{i,j} \left(\frac{E_\gamma}{\hbar c} \right)^5 \delta(E-E_i) 
\left|
\langle u_\ell^{(i)}(E_i,r) |r^2 | u_{\ell'}^{(j)}(E^\prime_j,r) \rangle
\right|^2
\label{cross2}
\end{equation}

In practice, the total cross section above is computed by making use of Eq.(\ref{deltas}) and replacing the 
$\delta$-function by $\delta_{E,E_i}/\Delta E$, which by use of Eq.(\ref{deltae}), 
permits finally to write the total cross section Eq.(\ref{cross2}) as:
\begin{equation}
\sigma(E_i)= \frac{4\pi^2 e^2}{15 k_i^2} (2\ell+1)  
\langle \ell 0; 2 0| \ell^\prime 0\rangle^2 \frac{L}{2\pi} \frac{k_i}{E_i}
\sum_{j} \left(\frac{E_\gamma}{\hbar c} \right)^5  
\left|
\langle u_\ell^{(i)}(E_i,r) |r^2 | u_{\ell'}^{(j)}(E^\prime_j,r) \rangle
\right|^2
\label{totdisc}
\end{equation}
with $E_\gamma=E_i-E^\prime_j$.  Since $E_\gamma>0$ it is then obvious that the summation in $j$ 
involves all the continuum final states with energy $E^\prime_j$ smaller than the energy $E_i$ of 
the initial state. 

Also, in the same way that the integration Eq.(\ref{intcs}) is restricted to final energies within 
a chosen final energy window, in Eq.(\ref{totdisc}) the summation over $j$ is also restricted to those
discrete final states whose energy $E'_j$ is contained is that energy window. It is important to note 
that to reach a sufficient accuracy in the calculation it is necessary to have a significant amount of 
discrete final energies within that window. 
This is because the relation Eq.(\ref{deltas}), thoroughly used in the description of the 
discretized continuum procedure, requires a sufficiently small value of $\Delta E$ (or equivalently,
a sufficiently large value of the box size $L$) to be valid. In fact, the summation over $j$
in Eq.(\ref{totdisc}) is actually taking care in the discretized picture of the integral 
Eq.(\ref{intcs}), and it is reasonable to think that too few terms in this summation can not reproduce
properly the value of the integral.

Another issue to note is that, as it has to be, the total cross
section Eq.(\ref{totdisc}) is independent of the size of the box $L$,
of course provided $L$ is sufficiently large. This is because the
number of states in the summation over $j$ increases linearly with
$L$, which together with the $L$ factor that explicitly appears in
Eq.(\ref{totdisc}) gives a total $L^2$-dependence.  This $L^2$-factor
cancels with the $1/L^2$ dependence of the square of the matrix
element in Eq.(\ref{totdisc}).  This $1/L^2$ dependence is evident
from the normalization of the asymptotic wave function
Eq.(\ref{asymp2}).

As a final remark, let us mention that Eq.(\ref{cross2}), and therefore also Eq.(\ref{totdisc}), is consistent
with the standard expression for the $\gamma$-decay cross section, as given for instance in Eq.(4) of
Ref.~\cite{for03}. The details about this consistency are given in appendix \ref{app2}.

\subsection{Discrete continuum states in a box: $E2$-capture in $\alpha+\alpha$ collisions.}

As done in the previous section, we shall now compute the $E2$-bremsstrahlung cross sections in 
$\alpha+\alpha$ collisions using the procedure described above.

As already mentioned, use of Eq.(\ref{totdisc}) requires a sufficiently small energy separation 
$\Delta E$ between the discrete continuum states, such that the number of terms involved in the 
summation over $j$ is large enough to reproduce the correct value of the integral Eq.(\ref{intcs}).
Typically, for a reasonably smooth function, about $20-30$ terms can be taken as a
lower limit for the number of states to be included in the summation.
Therefore, given a final energy window, use of Eq.(\ref{deltae}) permits to estimate
the size of the box $L$ needed for a reasonable description of the process.

For instance, for the $4^+ \rightarrow 2^+$ transition considered in Fig.\ref{fig2}b, where the final
energy window is $2 \mbox{ MeV} < E' < 4 \mbox{ MeV}$, an energy separation between 
states of about 0.1 MeV would give rise to around 20 discrete states within the window. 
According to Eq.(\ref{deltae}), for two 
alpha-particles and $E_i=3$ MeV, we get that in order obtain $\Delta E \sim 0.1$ MeV we need
the size of the box to be $L\sim 350$ fm. For the $2^+ \rightarrow 0^+$ reaction
(Fig.\ref{fig2}a) the energy window has a width of only 1 keV, which means that the required
separation energy between the discrete $0^+$ states should be of at most of about 0.1 keV. If we again
use Eq.(\ref{deltae}) with $E_i=0.1$ MeV ($\sim 0^+$ resonance energy) we obtain that to get
such a small energy separation we would need $L\sim 65000$ fm. This value would be even three
orders of magnitude bigger in case of looking for separation energies of the order of 0.1 eV, which
would be actually more reasonable, since the width of the $0^+$ resonance in $^8$Be is of just a
few eV. 

The huge size of the box estimated for the $2^+ \rightarrow 0^+$ transition is far beyond our numerical 
capability, and makes the discretization method completely useless in this particular case. 
This is due to the fact that the width of the window is orders of magnitude smaller than the energy
in the center of the window. Only when these two values are comparable we can say that the use of 
the discretization method really makes sense. For this reason, in this section we shall consider 
only the case of the $4^+ \rightarrow 2^+$ transition.

As before, the $4^+$ and $2^+$ wave functions have been obtained by
solving Eq.(\ref{schr}) with the Buck $\alpha$-$\alpha$ potential
\cite{buc77}. They have been computed by imposing a box boundary
condition, where normalization automatically is ensured.  However, the
matrix elements are still strongly oscillating, and the physical
meaning has to be extracted as by using the full continuum method as
described in Section~\ref{sec2}.  Again we use the Zel'dovich
prescription where each of the oscillating wave functions are
multiplied by the Zel'dovich factor, $e^{-\eta^2 r^2/2}$.  This
changes the normalization to be a function of $\eta$ and the matrix
elements should be correspondingly computed with this $\eta$-dependent
normalization. In the limit of $\eta \rightarrow 0$ the original
box-normalization is recovered and the radial integral is obtained
precisely as for continuum wave functions.  The convergence properties
and the range of acceptable parameters are therefore unchanged.  In
the numerical results we use the value $\eta=0.01$ fm$^{-1}$ for the
Zel'dovich parameter.  This implies that the minimum size of the box
should be of about $L=500$ fm, since this value of $L$ guarantees that
the integral in Eq.(\ref{totdisc}) has already converged (see inset of
Fig.\ref{fig1}).  Finally, the corresponding cross section has been
computed according to Eq.(\ref{totdisc}).

\begin{figure}
\centering
\includegraphics[width=10cm]{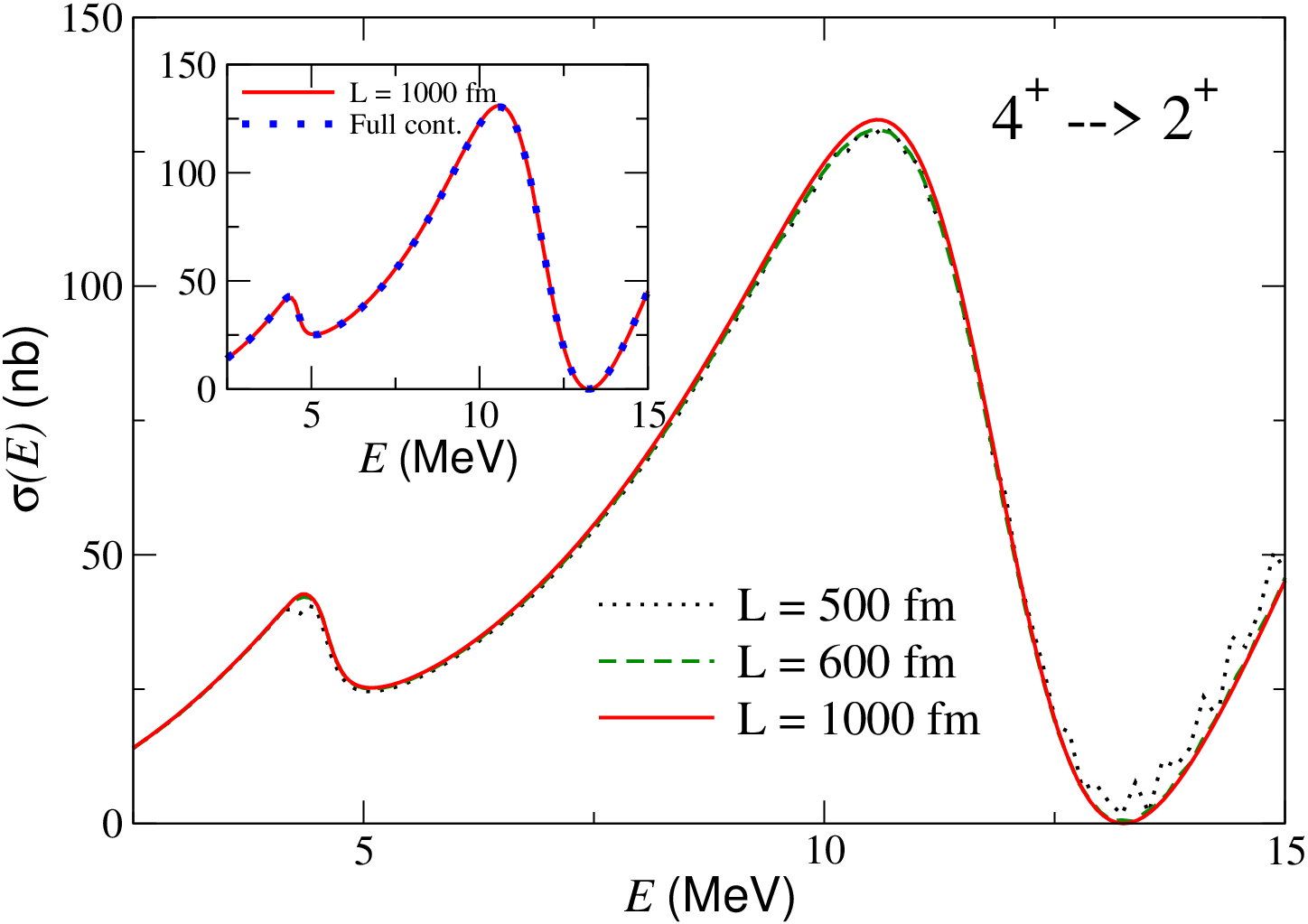}
\caption{The same as in Fig.~\ref{fig2}b but using continuum states discretized with a box boundary
condition. The dotted, dashed, and solid curves have been obtained with a box size
of $L=$ 500, 600, and 1000 fm, respectively. The inset compares the result obtained with 
$L=1000$ fm (solid curve) and the cross section shown in Fig.\ref{fig2}b obtained with the
full continuum calculation (dotted curve).  } 
\label{fig3}       
\end{figure}

To test the dependence on the size of the box, we have performed the calculation for different values of
$L$. In Fig.\ref{fig3} the dotted curve is the cross section obtained with $L=500$ fm. We can see that,
specially for high energies, the curve does not show a smooth behavior. This is due to numerical
inaccuracies produced by a still not large enough number of discrete $2^+$ states in the final 
energy window (30 states). In fact, if we reduce $L$ down to 400 fm (only 23 discrete  
states in the window), the non-smooth behavior is much more pronounced (although not shown
in the figure for the sake of clearness). On the other hand, an increase of $L$ up to
600 fm (36 discrete states within the window), permits already to get a smooth cross section,
as shown by the dashed curve in the figure. When increasing the value of $L$, and
therefore increasing as well the number of states inside the energy window, a small correction
is found for the cross section. This is shown by the solid curve, which has been obtained
with $L=1000$ fm (60 discrete $2^+$ states in the energy window). Further increase of $L$ does not
produce any visible change when compared to the solid curve in the figure. Finally, in the inset
we compare the converged cross section obtained with $L=1000$ fm (solid curve) and the one  
shown in Fig.\ref{fig2}b, obtained with the full continuum method, and plotted in 
the inset of Fig.\ref{fig3} by the dotted curve. As we can see the agreement is perfect.

\section{Complex Scaling Method}
\label{sec4}

It is well known that when the energy in Eq.(\ref{schr}) is allowed to
be complex, bound states and resonances can be identified as poles of
the $S$-matrix in the complex momentum plane.  In particular, bound
states are located in the positive side of the imaginary axis, and the
resonances appear in the fourth quadrant of the plane.  
The asymptotic behavior Eq.(\ref{asymp2}) for complex values of the
momentum $k$ implies that while the bound states are falling off
exponentially at large distances, the resonance wave functions do
actually diverge also exponentially.

The numerical difficulties arising from this divergence can however be easily solved by use of the complex
scaling method \cite{ho83,moi98}. Its application requires only rotation of the radial 
coordinate into the complex plane by some arbitrary angle $\theta$ ($r \rightarrow r e^{i\theta}$). Under
this simple transformation, and provided that $\theta$ is larger than the argument of the resonance, the
complex rotated resonance wave function behaves asymptotically as a bound
state, i.e., it falls off exponentially. The same behavior is maintained for bound states.
Therefore, after complex scaling, resonances appear formally as ``bound states'' with complex energy.

A complex scaling transformation permits then an easy distinction
between continuum states, which are rotated in the complex energy
plane by an angle $2\theta$ \cite{ho83}, and resonances, which show up
as isolated points whose position is independent of the complex
scaling angle used in the calculation. Therefore, this procedure appears as a 
simple tool allowing separation between different types of contributions: 
Resonance to resonance, continuum to resonance, resonance to continuum, or
continuum to continuum.

In the following, we shall discuss the general properties of the method and its applicability
to describe bremsstrahlung processes. Again, we shall apply the method to our
test case of two alpha-particles.

\subsection{General formulation }

\begin{figure}
\centering
\includegraphics[width=10cm]{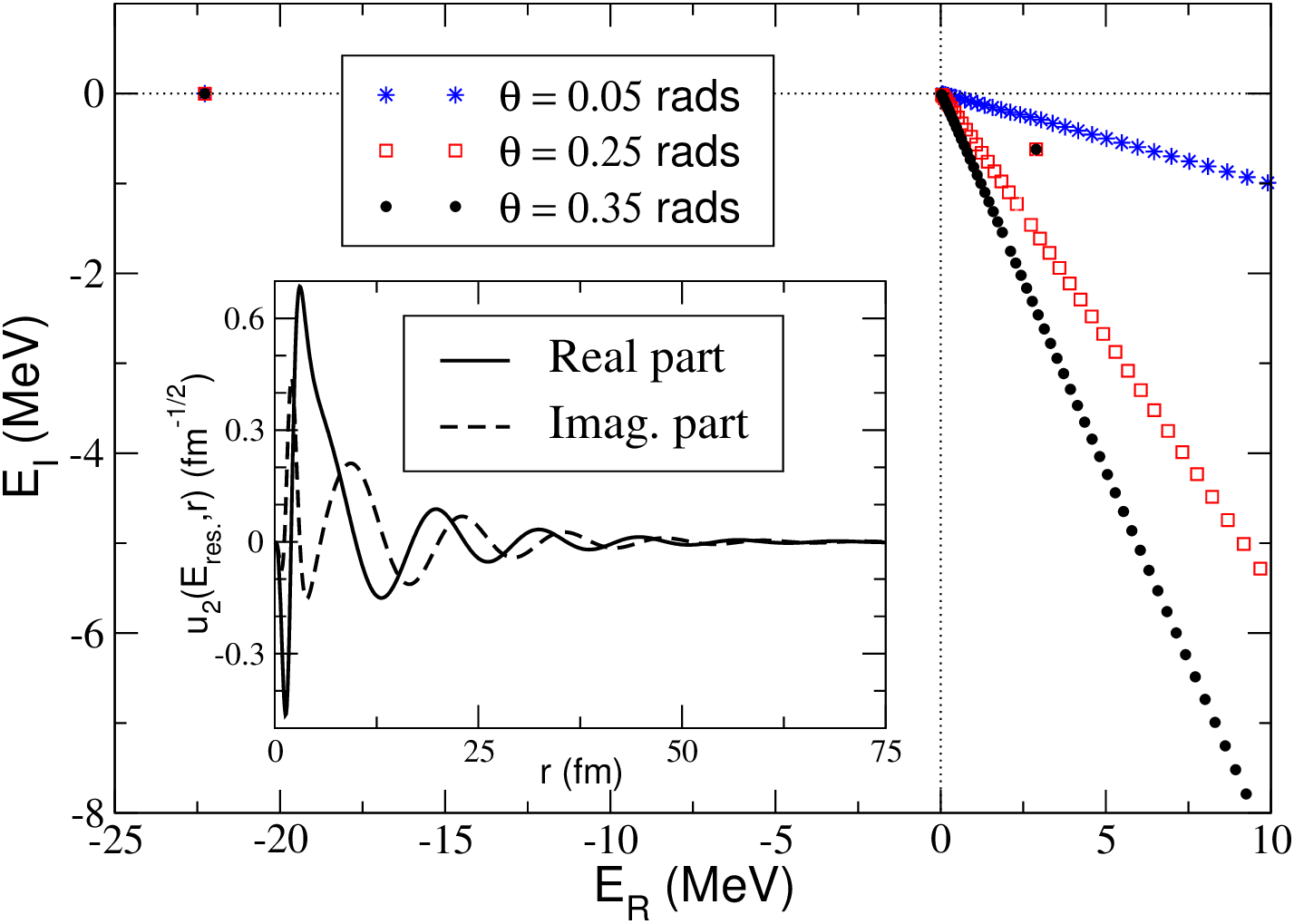}
\caption{Outer part: Complex rotated spectrum, in the complex energy plane, of the 
2$^+$ states in $^8$Be after solving the complex rotated Schr\"{o}dinger equation (\ref{schr}) with
the Buck potential \cite{buc77}, and a box boundary condition. The spectrum is shown for
scaling angles $\theta=0.05$ rads. (stars), $\theta=0.25$ rads. (squares), and 
$\theta=0.35$ rads. (circles). Inner part: The corresponding complex rotated radial wave function
(with $\theta=0.25$ rads.) for to the $2^+$ resonance in $^8$Be.} 
\label{fig4} 
\end{figure}

As an illustration of the discussion above, we show in
the outer part of Fig.~\ref{fig4} the $2^+$ spectrum in $^8$Be after
solving the complex rotated Schr\"{o}dinger equation Eq.(\ref{schr})
with the Buck potential \cite{buc77}, and imposing a box boundary
condition to the solutions. The results for three different scaling
angles (0.05, 0.25, and 0.35 rads.) are shown by the stars, squares, and
circles, respectively.  As seen in the figure,
the spurious Pauli forbidden bound 2$^+$ state at about $-22.5$ MeV is
found in all the three calculations. The $2^+$ resonance at the
complex energy $2.9-i0.6$ MeV is found as an isolated pole independent
of the scaling angle, but only for the two cases where the scaling angle is
larger than 0.10 rads, which is the argument of the resonance.  The
remaining points correspond to the original real energies of continuum
$2^+$ states rotated in the complex energy plane by an angle
$2\theta$.  In the case of $\theta=0.05$ rads (stars in the figure),
the $2^+$ resonance is not explicitly found, and its effect on the
cross section is then distributed among all the discrete continuum
states, as in section \ref{sec3}.

The inner part of the figure shows the complex rotated radial wave function of the
$2^+$ resonance in $^8$Be for a complex scaling angle of 0.25 rads. As we can see, for a distance of 75 fm
the radial wave function is already pretty small. It is then evident from the figure that after a complex
scaling transformation, the integral in Eq.(\ref{radial}) does not diverge anymore, provided that
at least one of the states involved in the calculation is a resonance (or a bound state). 
For transitions between pure continuum states the divergence problem remains, 
and some regularization (like the Zel'dovich regularization used in this work) would still be necessary.

In order to compute the bremsstrahlung cross section it is then very tempting to make a complex
scaling transformation and impose a box boundary condition, which discretizes the spectrum as
shown in Fig.~\ref{fig4}. Doing like this, it is also tempting to use the complex scaled version of
the method shown in Section~\ref{sec3}, which is based on Eqs.(\ref{deltae}) and (\ref{deltas}),
which give the energy separation between the different discrete continuum states and the relation
between the Dirac and Kronecker deltas, respectively.  In this way the cross section could again 
be computed as given in Eq.(\ref{totdisc}), but where the radial wave functions and the $r^2$ operator
have been complex rotated.

However, as shown in Ref.\cite{alo92}, the density of states in the complex rotated continuum does
not show the expected behavior. In fact, the simple expression Eq.(\ref{discr}), and therefore also
Eq.(\ref{deltae}), is found to be valid only in the limit of large rotation angle and large size of
the box. This is related to the fact that after complex scaling the resonance states are separated from the
other states in the continuum, and therefore one may expect to observe holes in the density of states in
the rotating continuum \cite{alo92} (as actually seen in Fig.\ref{fig4} for the continuum states shown
by the squares and the circles, which correspond to complex scaling angles such that the resonance is
taken out from the continuum spectrum).  In other words, the fact that resonances are isolated in the
complex energy plane precludes the use of Eq.(\ref{deltae}) in order to transform Eq.(\ref{cross2})
into Eq.(\ref{totdisc}).

The alternative is the method described in \cite{myo98,kat06}, where the complex scaling method is used
to investigate the transition from the continuum into a bound state. The starting point is the 
particularization of the transition strength given in appendix \ref{app2}, Eq.(\ref{eqb2}), where the 
continuum spectra is assumed
to be discretized, to the case of a transition into a bound state. This particularization reads:
\begin{equation}
\frac{d{\cal B}^{(\lambda)}}{dE}(J' \rightarrow J) 
 = \sum_{i} \langle\Phi_{J'}^{(bound)}|\hat{\cal O}^\dag_\lambda|\Phi^i_J\rangle\langle\Phi^i_J
    |\hat{\cal O}_\lambda| \Phi_{J'}^{(bound)}\rangle \delta(E-E_i),
\label{eq26}
\end{equation}
where for simplicity we have suppressed the indices referring to the projections of the angular momenta,
the summation $i$ runs over all the discrete states whose wave function is given by $\Phi^i_{J}$ 
(the continuum states among them), and where $\Phi^{(bound)}_{J'}$ is the wave function of the bound state.

In appendix \ref{app3} we have summarized the Green's function formalism used in \cite{myo98}, which
permits to relate the transition strength Eq.(\ref{eq26}) to the imaginary part of the so called response
function given by Eq. (\ref{resp}). This connection is given by Eq.(\ref{str}), which for the case of 
discrete continuum states takes the form: 
\begin{equation}
\frac{d{\cal B}^{(\lambda)}}{dE}(J' \rightarrow J) 
 = -\frac{1}{\pi} Im \left[
\sum_i \frac{\langle\Phi_{J'}^{(bound)}|\hat{\cal O}^\dag_\lambda|\Phi^i_J\rangle\langle\Phi^i_J
    |\hat{\cal O}_\lambda| \Phi_{J'}^{(bound)}\rangle}{E-E_i}
\right].
\label{eq27}
\end{equation}

The calculation of Eq.(\ref{eq27}) is particularly simple when performing a complex scaling transformation, 
in such a way that the wave functions, the operator, and the eigenvalues in Eq.(\ref{eq27}) become complex 
quantities.
In fact, after complex scaling the discrete complex scaled states, for instance the ones shown in 
Fig.\ref{fig4}, still form a complete basis \cite{gir03}. 
The advantage of this method is that all the $E_i$-values are now complex
and the initial energy $E$ is still real. This means that the summation in Eq.(\ref{eq27}) can now be easily
made for all values of $E>0$ (only the energies of the bound states, if any, are still real, but negative).

Therefore, after complex scaling of the $\alpha-\alpha$ potential, and imposing a box boundary condition, 
one gets the family of discrete eigenvalues shown in Fig.\ref{fig4}, each of them associated to a complex
rotated wave function. These functions are the complex rotated version of Eqs.(\ref{eqb7}) and
(\ref{eqb8}). Using also the complex scaled version of the electromagnetic operator Eq.(\ref{eqb6})
one can compute the matrix elements in Eq.(\ref{eq27}) and thereby the transition strength and
the cross section.

In the case of  transitions from continuum to continuum the response function Eq.(\ref{resp}) has to be 
written as:
\begin{equation}
{\cal R}_\lambda(E,E')=\sum_j \frac{1}{E'-E'_j}
\sum_i \frac{\langle\Phi_{J'}^{(j)}(E'_j)|\hat{\cal O}^\dag_\lambda|\Phi^i_J(E_i)\rangle
   \langle\Phi^i_J(E_i)|\hat{\cal O}_\lambda| \Phi_{J'}^{(j)}(E'_j)\rangle}{E-E_i},
\label{eep}
\end{equation}
where $i$ and $j$ run over the initial and final (discrete) continuum spectrum,  where we have
written explicitly the dependence of each wave function on the (complex) discrete energies 
$E_i$ and $E'_j$, and where $E$ and $E'$ are the (real) initial and final energies.

Making use now twice of Eq.(\ref{pv}) we get the analogous to Eq.(\ref{eq27}) for continuum to continuum
transitions:
\begin{eqnarray}
\lefteqn{
{\cal R}_\lambda(E,E')= } \nonumber \\ & &
P.V.\left[ \sum_j \frac{1}{E'-E'_j} P.V.\left[
\sum_i \frac{\langle\Phi_{J'}^{(j)}(E'_j)|\hat{\cal O}^\dag_\lambda|\Phi^i_J(E_i)\rangle
   \langle\Phi^i_J(E_i)|\hat{\cal O}_\lambda| \Phi_{J'}^{(j)}(E'_j)\rangle}{E-E_i}
\right] \right] - 
\nonumber \\ &&
-\pi^2 \sum_{ij} \langle\Phi_{J'}^{(j)}(E'_j)|\hat{\cal O}^\dag_\lambda|\Phi^i_J(E_i)\rangle
   \langle\Phi^i_J(E_i)|\hat{\cal O}_\lambda| \Phi_{J'}^{(j)}(E'_j)\rangle
  \delta(E-E_i) \delta(E'-E'_j)
\label{ctoc} \\ &&
- i \pi P.V. \left[ \sum_{j} 
\frac{\sum_i \langle\Phi_{J'}^{(j)}(E'_j)|\hat{\cal O}^\dag_\lambda|\Phi^i_J(E_i)\rangle
   \langle\Phi^i_J(E_i)|\hat{\cal O}_\lambda| \Phi_{J'}^{(j)}(E'_j)\rangle \delta(E-E_i)}{E'-E'_j}
 \right]
\nonumber \\ &&
- i \pi \sum_{j} P.V. \left[\sum_i
\frac{\langle\Phi_{J'}^{(j)}(E'_j)|\hat{\cal O}^\dag_\lambda|\Phi^i_J(E_i)\rangle
   \langle\Phi^i_J(E_i)|\hat{\cal O}_\lambda| \Phi_{J'}^{(j)}(E'_j)\rangle}{E-E_i} \right]
\delta(E'-E'_j) ,
\nonumber
\end{eqnarray}
where $P.V.$ means the Principal Value of the corresponding integral.

As we can see from the expression above, the transition strength Eq.(\ref{eqb2}) is contained 
in the real part of the response function Eq.(\ref{eep}).  Unfortunately,
this real part is contaminated by a double Principal Value, which is also real.
Use of a complex scaling transformation permits an easy calculation of the response function Eq.(\ref{eep}), 
and in particular of its real part. However, to eliminate
from it the double principal value given in Eq.(\ref{ctoc}) can be a rather complicated task. 
For this reason, the use of the complex scaling
method to compute bremsstrahlung cross sections appears to be complicate. The
exception can be those cases where a very well defined and narrow resonance is present in the final state.
This happens for instance in the $2^+ \rightarrow 0^+$ transition in $^8$Be, where the very low-lying
$0^+$ resonance is so narrow that it can be treated as a bound state, and therefore
Eq.(\ref{eq27}) can still be used.

\subsection{Complex scaling: $E2$-capture in the $2^+ \rightarrow 0^+$ transition in $^8$Be.}

We shall start this section by making the $s$-wave $\alpha$-$\alpha$ interaction (in principle
described by the Buck potential \cite{buc77}) slightly more attractive, such that the low-lying
0$^+$ resonance becomes a true bound state. Under these conditions, after a complex scaling
calculation of the initial and final states, Eq.(\ref{eq27}) can be used safely. With the radial
matrix elements computed in this way, the cross section is then obtained from Eq.(\ref{cross1}),
where $E_\gamma=E-E_B$, with $E_B$ being the binding energy of the artificially bound $0^+$ state. 

\begin{figure}
\centering
\includegraphics[width=10cm]{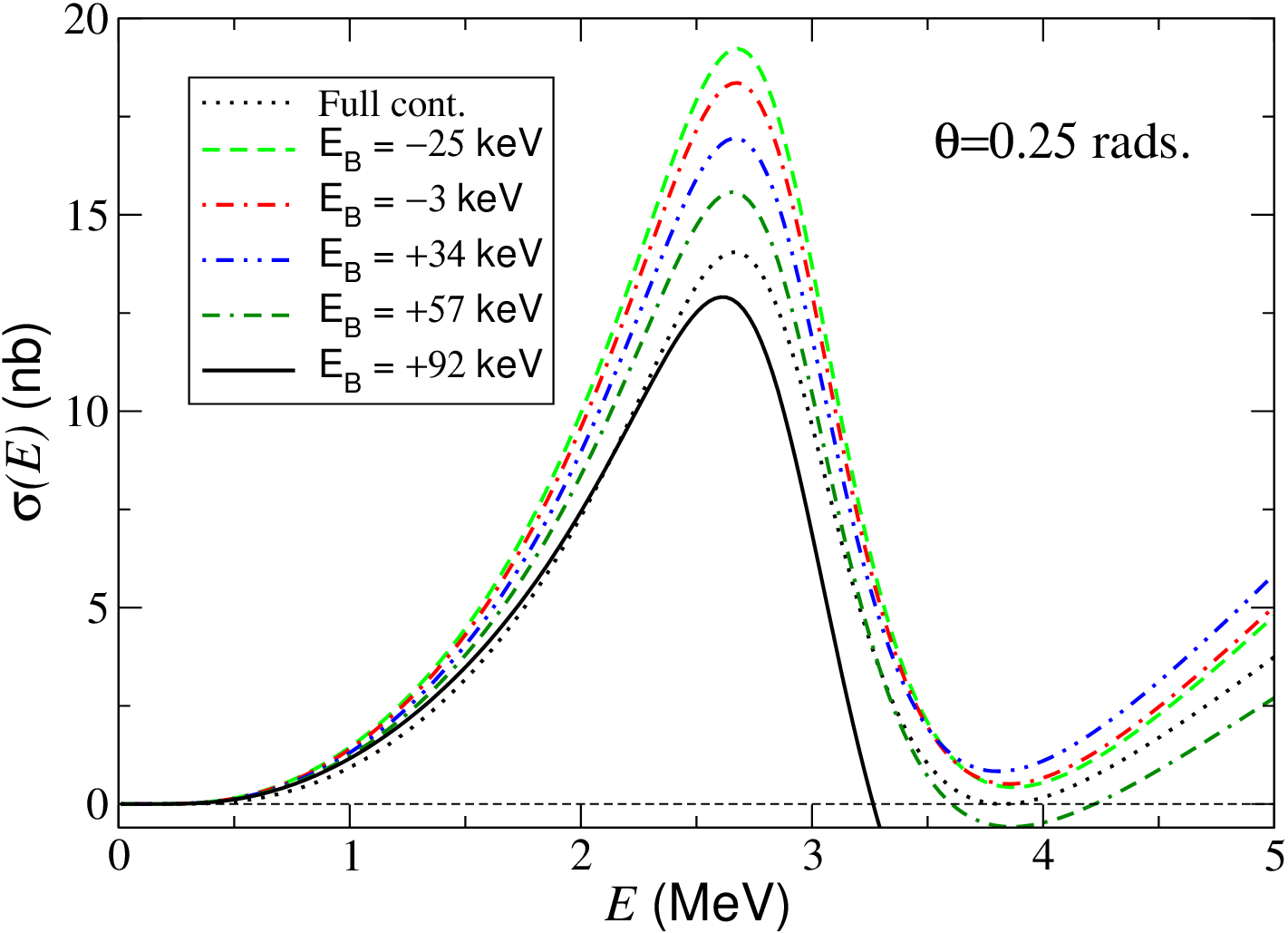}
\caption{Cross section for the $2^+ \rightarrow 0^+$ transition into the ground state of $^8$Be  
for different energies ($E_B$) of the $0^+$ state. The complex scaling calculation has been
made with a complex scaling angle of 0.25 rads. The result given in Fig.\ref{fig2}a (dotted curve)
is shown for comparison.}
\label{fig6}
\end{figure}

The dashed line in Fig.\ref{fig6} shows the computed cross section when the $0^+$ state has
a binding energy of $E_B=-25$ keV. The calculation has been done with a complex scaling angle 
$\theta=0.25$ rads., but the result is of course independent of the angle used.  For comparison, 
we also show in the figure the cross section given in Fig.{\ref{fig2}a (dotted curve),
corresponding to a calculation on the real energy axis with the true Buck potential. 
As we can see, the increase in the $0^+$ binding energy produces
an increase in the cross section, with a maximum value that goes up from about 14 nb to about 19 nb.
In fact, for a smaller binding like $E_B=-3$ keV (dot-dashed curve in the figure) the maximum
of the cross section goes down to 18 nb.

If we still reduce the attraction in the $s$-wave $\alpha$-$\alpha$ potential the bound $0^+$
state becomes a true resonance. Still using Eq.(\ref{eq27}), we then get the cross sections
shown by the dot-dot-dashed curve, which corresponds to a $0^+$ resonance energy of 34 keV, and
the dashed-dashed-dot curve, which corresponds to a $0^+$ resonance energy of 57 keV. 
As we can see, the more the resonance energy approaches the experimental value of 92 keV, the more the
cross section approaches the result obtained on the real energy axis. However, for a resonance 
energy of 57 keV, we observe that at some point, in the vicinity of $E=4$ MeV, the cross section 
is negative, which is already an indication that use of Eq.(\ref{eq27}) is not fully correct
for all energies.
First, the energy of the final state is not fully real, and the square of the radial matrix
element Eq.(\ref{radial}) would be in general complex. Therefore it can not be the real quantity 
given by Eq.(\ref{eq27}). And second, the contributions not included in the calculation (transitions to 
the continuum 0$^+$ states) have
already some role to play. In fact, when the correct Buck potential is used for the $s$-wave 
interaction (0$^+$ resonance at 92 keV), the cross section obtained from Eq.(\ref{eq27}), solid 
curve in the figure, becomes very negative for initial energies higher than about 3 MeV. Also,
the maximum of the cross section is 1.5 nb below the result shown in Fig.\ref{fig2}a.

\begin{figure}
\centering
\includegraphics[width=10cm]{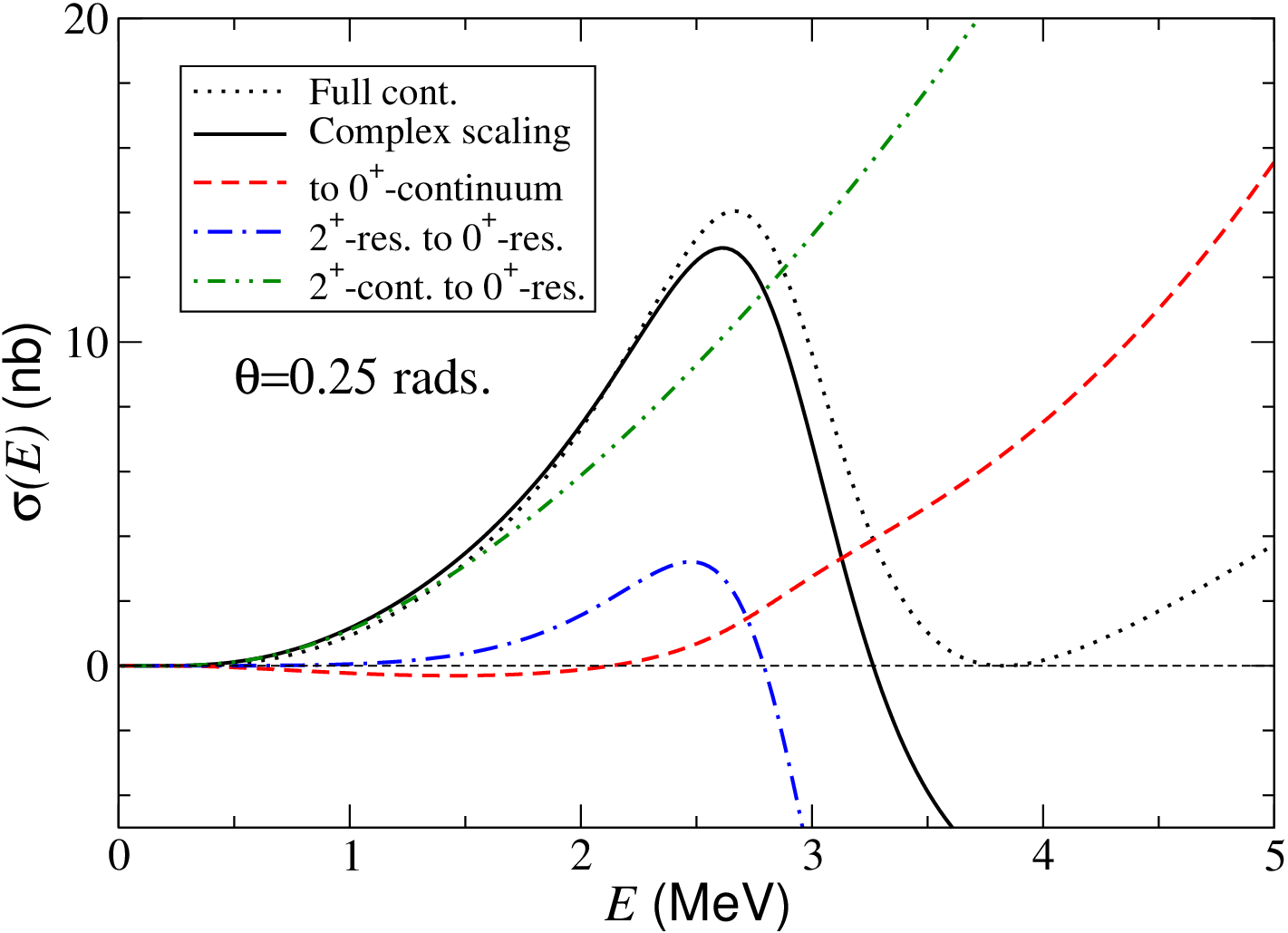}
\caption{Contributions of the different transitions to the bremsstrahlung cross section for the 
$2^+ \rightarrow 0^+$ transition in $^8$Be. The scaling angle $\theta$ has been
taken equal to 0.25 rads. The solid line gives the total cross section obtained from Eq.(\ref{eq27})
assuming transitions to the $0^+$ resonance only (which is treated as a bound state). The dotted line 
is the cross section obtained
with a full continuum calculation on the real energy axis (Fig.\ref{fig2}a). The difference between 
these two curves is given by the dashed curve. The contributions from the resonance-resonance and
continuum-resonance transitions given by Eq.(\ref{eq27}) are shown by the dot-dashed and the
dot-dot-dashed curves, respectively.  }
\label{fig7}
\end{figure}

In any case, even if Eq.(\ref{eq27}) does not contain the full information about the 
cross section, it can be used to extract some of the contributions to it. In particular, the
summation in the r.h.s. of the equation permits to separate the contribution of transitions
from the continuum $2^+$ states to the resonant $0^+$ state (continuum-resonance contribution), 
and the contribution corresponding to a resonance-resonance transition. Of course, for this
separation to be possible, a complex scaling angle larger than the argument of the $2^+$ resonance
is required. This is already indicating that this separation between different types of contributions
depends on the complex scaling angle.

The different contributions are shown in Fig.\ref{fig7}, where we plot again the solid and dotted curves
already shown in Fig.\ref{fig6}, which correspond to the cross section obtained from Eq.(\ref{eq27})
assuming transitions to the 0$^+$ resonance only (and treated as a bound state), and the full cross 
section computed on the real energy axis. The difference between these two curves (dashed curve in 
the figure) represents the effect of, first, the transitions to continuum $0^+$ states not included in 
Eq.(\ref{eq27}), and, second, the fact that Eq.(\ref{eq27}) is not really valid for unbound final states 
(Eq.(\ref{ctoc}) should be used instead). Due to very small width of the $0^+$ resonance, the effect
shown by the dashed curve in Fig.\ref{fig7} is expected to come mainly from the missing transitions
to the continuum $0^+$ states. The resonance-resonance and continuum-resonance contributions
obtained from Eq.(\ref{eq27}) are shown by the dot-dashed and dot-dot-dashed curves, respectively.
As we can see, the resonance-resonance contribution shows a peak at about the 2$^+$ resonance energy, but
right after the peak goes sharply down and it actually becomes very negative. This very negative 
contribution has to be compensated by the remaining ones, especially by the continuum-resonance 
contribution. However, even this is not enough, and the transitions to continuum final states play an
important role.

\section{Summary and conclusions}
\label{sec7}

In this work we have revisited the problem of bremsstrahlung radiation in two-body collisions, which
is equivalent to a gamma decay process between continuum states. The case of the $E2$ transitions 
in $\alpha$+$\alpha$ collisions is taken as an example, and used to test the methods. 

We have given the details, related, and compared three different procedures. In two of them the 
energies of the initial and final states are kept on the real energy axis. The main difference between 
them is in the way how the continuum states are treated. In the first method they are obtained with the 
correct asymptotic behavior, which implies an orthogonality between states in the continuum sense, with 
a Dirac delta. In the second method the continuum is discretized by imposing a box boundary condition. 
The continuum states are zero outside the box, and they are treated as bound states, and therefore 
normalized to 1 inside the box. Both procedures are of course consistent, and they have been shown 
to be fully equivalent in the limit of an infinitely big box. However, when very narrow resonances
are involved in the reaction under investigation, a correct description of such resonance with the second
method requires a huge discretization box, making this second procedure impractical for these cases
(like the $2^+\rightarrow 0^+$ process in $^8$Be). Although formally it is not needed,
the numerical implementation of these two methods is very much simplified after regularization of the
radial integrals involved in the calculation. In this work the Zel'dovich prescription has been used.

The third procedure is based on the complex scaling method. After complex scaling the resonances
appear formally as bound states with complex energy, and they are then well identified and differentiated
from ordinary continuum states. Once this is done, it is then, at least in principle, possible to 
separate the contribution from the different types of transition, namely, from resonance to resonance,
from continuum states to resonance, from resonance to continuum, and from continuum to continuum. 
However, although the method works well for transitions into bound states, we have shown that when
the final state is unbound the extraction of the radial integrals requires knowledge of a double
Principal Value, which makes the whole procedure not easy to implement. For sufficiently
low-lying narrow resonances in the final state the complex scaling procedure still provides a rather 
good approximation of the bremsstrahlung radiation cross section, making possible to estimate
the different contributions. In any case, even if the separation into different contributions
is made, only the sum of all of them is an observable physics quantity.

\begin{acknowledgements}
This work was partly supported by funds provided by DGI of MINECO (Spain) under contract No. FIS2011-23565.
\end{acknowledgements}

\appendix

\section{Energy normalization of the continuum wave functions}
\label{app1}
Let us start with a partial wave expansion of a two-body wave function:
\begin{equation}
\Psi(\bm{k},\bm{r})=\frac{1}{C} \sqrt{\frac{2}{\pi}}\frac{1}{kr} 
\sum_\ell i^\ell u_\ell(k,r) \sum_m Y_{\ell m}(\Omega_r) Y_{\ell m}^*(\Omega_k).
\end{equation}
In the expression above the constant $C$ can in principle be anything. The only requirement is that
the radial wave functions $u_\ell$ have to be normalized in such a way that $\Psi$ reduces
to a plane wave in the limit of no interaction between particles. In other words, $\Psi$ has to fulfill that:
\begin{eqnarray}
\lefteqn{
\Psi(\bm{k},\bm{r}) \stackrel{\mbox{\scriptsize free case}}{\rightarrow} 
\frac{1}{(2\pi)^{3/2}} e^{i \bm{k}\cdot\bm{r}} }
\nonumber \\ &&
=\sqrt{\frac{2}{\pi}} \sum_\ell i^\ell j_\ell(kr) \sum_m Y_{\ell m}(\Omega_r) Y_{\ell m}^*(\Omega_k),
\end{eqnarray}
from where it is obvious that $u_\ell$ must satisfy:
\begin{equation}
 u_\ell(k,r)
\stackrel{\mbox{\scriptsize free case}}{\rightarrow}
C kr j_\ell(kr).
\end{equation}

If we now make use of the fact that:
\begin{equation}
\int_0^\infty kr j_\ell(kr) k^\prime r j_\ell(k^\prime r) dr= \frac{\pi}{2}\delta(k-k^\prime),
\end{equation}
we then immediately see that the radial wave functions $u_\ell$ satisfy the normalization condition:
\begin{equation}
\int_0^\infty u_\ell(k,r) u_\ell(k^\prime,r) dr= C^2 \frac{\pi}{2} \delta(k-k^\prime).
\end{equation}
Therefore, if we choose $C=\sqrt{2/\pi}$ the radial wave functions would satisfy the so-called
momentum normalization.

Since $k=\sqrt{2\mu E/\hbar^2}$, it is not difficult to see that:
\begin{equation}
\delta(k-k^\prime)=
\delta(\sqrt{\frac{2\mu E}{\hbar^2}}-\sqrt{\frac{2\mu E^\prime}{\hbar^2}})
=\frac{\hbar^2 k}{\mu} \delta(E-E^\prime), 
\end{equation}
which leads to:
\begin{equation}
\int_0^\infty u_\ell(k,r) u_\ell(k^\prime,r) dr=  
C^2 \frac{\pi}{2} \frac{\hbar^2 k}{\mu} \delta(E-E^\prime),
\end{equation}
which implies that, as given in Eq.(\ref{constant}), by choosing
\begin{equation}
C=\sqrt{\frac{2\mu}{\pi \hbar^2 k}}
\end{equation}
the energy normalization condition
\begin{equation}
\int_0^\infty u_\ell(k,r) u_\ell(k^\prime,r) dr= \delta(E-E^\prime)
\end{equation}
is then satisfied.

\section{Radiative capture cross section.}
\label{app2}

Let us consider the photodissociation reaction $A+\gamma \rightarrow a+b$, and let us denote by
$J'$ and $J$ the angular momenta of the system $A$ and the continuum two-body system $ab$, respectively.
The corresponding photodissociation cross section (with multipolarity $\lambda$) can be found
in Ref.\cite{for03} for the case in which $A$ represents a bound state. When the state $A$ corresponds
also to a two-body continuum state (made of particles $a$ and $b$) the expression in Ref.\cite{for03}
can be generalized to:
\begin{equation}
\frac{d\sigma_\gamma^{(\lambda)}}{dE'}(E)=
\frac{(2 \pi)^3(\lambda+1)}{\lambda \left[ (2\lambda+1)!!\right]^2}
\left( \frac{E_\gamma}{\hbar c} \right)^{2\lambda-1} \frac{d{\cal B}^{(\lambda)}}{dE dE'}(J'\rightarrow J),
\label{eqb1}
\end{equation}
where $E$ and $E'$ are the energies of the final and initial states, and $E_\gamma$ is the photon 
energy.

We shall assume the initial and final continuum spectra to be discretized, in such a way that the 
transition strength $d{\cal B}^{(\lambda)}/dEdE'$ can be written as \cite{for03}:
\begin{eqnarray}
\frac{d{\cal B}^{(\lambda)}}{dEdE'}(J' \rightarrow J)& = & \frac{1}{2J'+1}
\sum_{i,j} |\langle \Phi^i_{J} || \hat{O}_{\lambda} ||\Phi^j_{J'} \rangle |^2 \delta(E-E_i) \delta(E'-E'_j)
\nonumber \\ & = &\sum_{i,j} \sum_{m, \mu}
|\langle \Phi^i_{J\mu} | \hat{O}_{\lambda m} |\Phi^j_{J'\mu'} \rangle |^2 \delta(E-E_i) \delta(E'-E'_j),
\label{eqb2}
\end{eqnarray}
where $|\Phi^j_{J'\mu'}\rangle$ and  $|\Phi^i_{J\mu}\rangle$ are the wave functions describing
the continuum state $A$ (with energy $E'_j$ and angular momentum and projection $J'\mu'$), and 
of the continuum final $ab$-system (with energy $E_i$ and angular momentum and projection $J\mu$), 
respectively.  The operator $\hat{O}_{\lambda}$ is the electromagnetic transition operator 
with rank $\lambda$, and the indices $i$ and $j$ run over all the (discrete) initial and 
final continuum states.

From this expression it is easy to connect the transition strength for a given reaction and the inverse one:
\begin{equation}
\frac{d{\cal B}^{(\lambda)}}{dEdE'} (J' \rightarrow J)  =\frac{2J+1}{2J'+1}
\frac{d{\cal B}^{(\lambda)}}{dEdE'} (J \rightarrow J').
\label{eq42}
\end{equation}

The photoabsorption cross section in Eq.(\ref{eqb1}) and the one corresponding to the inverse process, 
i.e., the radiative capture cross section $\sigma^{(\lambda)}(E)$ for the process 
$a+b \rightarrow A+\gamma$, are related by the detailed balance principle, which is given in Eq.(3) 
of \cite{for03}:
\begin{equation}
\frac{d\sigma^{(\lambda)}}{dE'}(E)=\frac{2(2J'+1)}{(2J_a+1)(2J_b+1)} \frac{1}{k^2}
          \left( \frac{E_\gamma}{\hbar c} \right)^2 \frac{d\sigma^{(\lambda)}_\gamma}{dE'}(E), 
\end{equation}
where $J_a$ and $J_b$ are the angular momenta of particles $a$ and $b$, respectively, 
and  $k^2=2\mu_{ab}E/\hbar^2$. Thanks to this 
relation, and making use of Eqs.(\ref{eqb1}) and (\ref{eq42}) we find the following general expression 
for the radiative capture cross section $a+b\rightarrow A+\gamma$:
\begin{equation}
\frac{d\sigma^{(\lambda)}}{dE'}(E)=\frac{(2 \pi)^3(\lambda+1)}{\lambda \left[ (2\lambda+1)!!\right]^2}
\frac{1}{k^2} 
\frac{2(2J+1)}{(2J_a+1)(2J_b+1)} 
\left( \frac{E_\gamma}{\hbar c} \right)^{2\lambda+1} 
\frac{d{\cal B}^{(\lambda)}}{dEdE'}(J\rightarrow J').
\label{eqb5}
\end{equation}

Let us consider now that particles $a$ and $b$ are identical, with spin zero, and with charge
$Ze$ (with $e$ the electron charge).  For the case of an electric transition process of 
order $\lambda$ the electromagnetic transition operator reads:
\begin{equation}
\hat{O}_{\lambda m}=e\sum_{n=1}^{2}Z_n r_n^\lambda Y_{\lambda m}(\Omega_r)=
\frac{Ze}{2^{\lambda-1}} r^\lambda Y_{\lambda m}(\Omega_r),
\label{eqb6} 
\end{equation}
where $\bm{r}=\bm{r}_1-\bm{r}_2$ and $\bm{r}_n$ is the center of mass coordinate of particle $n$.

Assuming now a central interaction between the two particles, the two-body wave functions involved
in Eq.(\ref{eqb2}) can be written as: 
\begin{eqnarray}
\Phi^j_{J'\mu'}(\bm{r})&=&\frac{u^{(j)}_{J'}(E'_j,r)}{r}Y_{J'\mu'}(\Omega_r)  \label{eqb7} \\
\Phi^i_{J\mu}(\bm{r})&=&\frac{u^{(i)}_{J}(E_i,r)}{r}Y_{J\mu}(\Omega_r)  \label{eqb8},
\end{eqnarray}
Inserting the expressions above and Eq.(\ref{eqb6}) into Eq.(\ref{eqb2}), and after analytical 
integration over the angular coordinates, we get that:
\begin{equation}
\frac{d{\cal B}^{(\lambda)}}{dEdE'}(J \rightarrow J') =
\frac{(Ze)^2}{2^{2\lambda-2}} \frac{1}{4\pi} \sum_{i,j} \delta(E-E_i) \delta(E'-E'_j)(2\lambda+1)
\left|
\langle J 0; \lambda 0 | J' 0 \rangle
\int dr u^{(i)}_{J}(E_i,r) r^\lambda u^{(j)}_{J'}(E'_j,r)
\right|^2,
\label{eqb9}
\end{equation}
from which Eq.(\ref{eqb5}), corresponding to the $E\lambda$ radiative capture process $a+b\rightarrow A+\gamma$,
takes the final form:
\begin{eqnarray}
\frac{d\sigma^{(\lambda)}}{dE'}(E)&=&
\frac{(Ze)^2}{2^{2\lambda-2}} 
\frac{2 \pi^2(\lambda+1)}{\lambda \left[ (2\lambda+1)!!\right]^2}
\frac{1}{k^2} 
\frac{2(2J+1)}{(2J_a+1)(2J_b+1)} 
\nonumber \\ & \times &
\sum_{i,j} 
\left( \frac{E_\gamma}{\hbar c} \right)^{2\lambda+1} 
\delta(E-E_i) \delta(E'-E'_j)(2\lambda+1)
\left|
\langle J 0; \lambda 0 | J' 0 \rangle
\int dr u^{(i)}_{J}(E_i,r) r^\lambda u^{(j)}_{J'}(E'_j,r)
\right|^2.
\end{eqnarray}

In the particular case of two $\alpha$ particles ($J_a=J_b=0$, $Z=2$) and an electric 
quadrupole transition ($\lambda=2$), and after integration over $E'$, we get for the total
cross section:
\begin{equation}
\sigma^{(\lambda)}(E)= \frac{2 \pi^2 e^2}{15 k^2} (2\ell+1) 
\langle \ell 0; 2 0 | \ell' 0 \rangle^2
\sum_{i,j}  \left( \frac{E_\gamma}{\hbar c} \right)^5 \delta(E-E_i)
\left|   \int dr u^{(i)}_{\ell}(E_i,r) r^2 u^{(j)}_{\ell'}(E'_j,r)  \right|^2,
\end{equation}
where we have replaced $J$ by $\ell$ and $J'$ by $\ell'$.

This result agrees with the one given in Eq.(\ref{cross2}), which in turns comes from the general
expression Eq.(\ref{cross1}), except for a factor of 2. This difference comes from the factor of 2
introduced in \cite{ald56,tan85} due to the fact that we are dealing with two identical particles (see
for instance Eq.(9) in Ref.~\cite{tan85}).

\section{Green's function formalism}
\label{app3}

In this section we summarize the aspects of the Green's function formalism that are relevant for
this work. All the details can be found in \cite{eco06}.

Let us consider a system whose hamiltonian operator is given by $\hat{\cal H}$, and such that its spectrum
is formed by a set of discrete states $\{|\Phi_n\rangle\}$ and the continuum states $|\Phi_c\rangle$.
The eigenfunctions form a complete basis, and the unity operator takes the form:
\begin{equation}
\mathbbm{1}=\sum_n |\Phi_n\rangle \langle \Phi_n | +\int dE_c |\Phi_c\rangle \langle \Phi_c |.
\label{unity}
\end{equation}

If $|\bm{r}\rangle$ denotes the eigenvector of the position operator, we have that 
$\langle \bm{r} |\bm{r}' \rangle=\delta (\bm{r}-\bm{r}')$, the unity operator can also be written
as 
\begin{equation}
\mathbbm{1}=\int d\bm{r} |\bm{r}\rangle \langle \bm{r}|,
\label{unib}
\end{equation}
and the Schr\"{o}dinger equation $\hat{\cal H}|\Phi_n\rangle = E_n |\Phi_n\rangle$ can be written
in coordinate space as:
\begin{equation}
\int d\bm{r} d\bm{r}' |\bm{r}\rangle \langle \bm{r}| \hat{\cal H} |
             \bm{r}'\rangle \langle \bm{r}'|\Phi_n\rangle =
             \int d\bm{r} E_n  |\bm{r}\rangle \langle \bm{r}| \Phi_n\rangle.
\end{equation}

The functions $\langle \bm{r}| \Phi_n\rangle=\Phi_n(\bm{r})$ are the eigenfunctions of the hamiltonian
in coordinate space, ${\cal H}(\bm{r}) \Phi_n(\bm{r})=E_n \Phi_n(\bm{r})$, is such a way the equation 
above can be written also as:
\begin{equation}
\int d\bm{r} d\bm{r}' |\bm{r}\rangle \langle \bm{r}| \hat{\cal H} |
             \bm{r}'\rangle \Phi_n(\bm{r}')  =
             \int d\bm{r} |\bm{r}\rangle {\cal H}(\bm{r})\Phi_n(\bm{r}),
\end{equation}
from which we can immediately get that:
\begin{equation}
\langle \bm{r}| \hat{\cal H} | \bm{r}'\rangle = {\cal H}(\bm{r}) \delta(\bm{r}-\bm{r}').
\label{hr}
\end{equation}

The Green's function for a given energy $E$ is defined as the function $G(E;\bm{r},\bm{r}')$ satisfying
that:
\begin{equation}
\left(E-{\cal H}(\bm{r})\right)G(E;\bm{r},\bm{r}') = \delta(\bm{r}-\bm{r}').
\label{def}
\end{equation}

The Green's function can also be defined in terms of the operator $\hat{G}(E)$, such that:
\begin{equation}
G(E;\bm{r},\bm{r}')=\langle \bm{r} | \hat{G}(E) | \bm{r}'\rangle.
\label{defg}
\end{equation}

The form of the $\hat{G}(E)$ operator can be obtained by noting that:
\begin{equation}
\langle \bm{r} | (E-\hat{\cal H})\hat{G}(E) | \bm{r}'\rangle =
\int d\bm{r}'' \langle \bm{r} | E-\hat{\cal H} | \bm{r}''\rangle \langle \bm{r}''|\hat{G}(E)|\bm{r}'\rangle,
\end{equation}
where we have made use of Eq.(\ref{unib}). Having now in mind Eqs.(\ref{hr}) and (\ref{defg}), the
expression above then reads:
\begin{equation}
\langle \bm{r} | (E-\hat{\cal H})\hat{G}(E) | \bm{r}'\rangle =
\int d\bm{r}'' \delta(\bm{r}-\bm{r}'') (E-{\cal H}(\bm{r}))G(E;\bm{r},\bm{r}'),
\end{equation}
which due to Eq.(\ref{def}) leads to:
\begin{equation}
\langle \bm{r} | (E-\hat{\cal H})\hat{G}(E) | \bm{r}'\rangle=\delta(\bm{r}-\bm{r}'), 
\end{equation}
or, in other words:
\begin{equation}
\hat{G}(E)=\frac{\mathbbm{1} }{E-\hat{\cal H}}.
\label{oper}
\end{equation}

Application of the operator (\ref{oper}) on the unity operator (\ref{unity}) permits to write:
\begin{equation}
\hat{G}(E)=\sum_n \frac{|\Phi_n\rangle \langle \Phi_n |}{E-E_n} +
\int dE_c \frac{|\Phi_c\rangle \langle \Phi_c |}{E-E_c},
\label{expan1}
\end{equation}
or, by the definition (\ref{defg}):
\begin{equation}
G(E;\bm{r},\bm{r}')=
\sum_n \frac{\Phi_n(\bm{r})  \Phi_n^*(\bm{r}') }{E-E_n} +
\int dE_c \frac{\Phi_c(\bm{r}) \Phi_c^*(\bm{r}') }{E-E_c}.
\end{equation}

Let us consider now some operator $\hat{\cal O}$, and let us consider the matrix element
$\langle \Phi_n | \hat{\cal O}^\dag \hat{G}(E) \hat{\cal O} | \Phi_n \rangle$, which thanks to
Eq.(\ref{expan1}) can be written as:
\begin{equation}
\langle \Phi_n | \hat{\cal O}^\dag \hat{G}(E) \hat{\cal O} | \Phi_n \rangle =
\sum_m \frac{\langle\Phi_n|\hat{\cal O}^\dag|\Phi_m\rangle\langle\Phi_m|\hat{\cal O}|\Phi_n\rangle}{E-E_m} +
\int dE_c\frac{\langle \Phi_n|\hat{\cal O}^\dag|\Phi_c\rangle\langle\Phi_c|\hat{\cal O}|\Phi_n\rangle}{E-E_c}.
\label{resp}
\end{equation}

If the hamiltonian $\hat{\cal H}$ is hermitian all the eigenvalues are real, and the function above
has a series of poles on the real energy axis. The expression above has to be computed making use
of:
\begin{eqnarray}
\lefteqn{
\lim_{y \rightarrow 0^+} \int_A^B \frac{f(x)}{x \pm iy}dx= } \nonumber \\ & &
\lim_{\alpha \rightarrow 0^+} \left[ 
\int_A^{-\alpha} \frac{f(x)}{x}dx + \int_\alpha^B \frac{f(x)}{x}dx
\right] \mp i \pi \int dx f(x) \delta(x)=
P.V. \left[ 
\int_A^B \frac{f(x)}{x}dx \right] \mp i \pi \int dx f(x) \delta(x),
\label{pv}
\end{eqnarray}
where $A<0<B$ and $f(x)$ is a function well behaved in the interval $[A,B]$, and where $P.V.$ refers 
to the Principal Value of the integral. Using (\ref{pv}) we can then obtain:
\begin{equation}
\sum_m \langle\Phi_n|\hat{\cal O}^\dag|\Phi_m\rangle\langle\Phi_m|\hat{\cal O}|\Phi_n\rangle \delta(E-E_m)
+ \int dE_c \langle \Phi_n|\hat{\cal O}^\dag|\Phi_c\rangle\langle\Phi_c|\hat{\cal O}|\Phi_n\rangle
          \delta(E-E_c) = -\frac{1}{\pi} Im\left(
\langle \Phi_n | \hat{\cal O}^\dag \hat{G}(E) \hat{\cal O} | \Phi_n \rangle
                                               \right).
\label{str}
\end{equation}

The function $\langle \Phi_n | \hat{\cal O}^\dag \hat{G}(E) \hat{\cal O} | \Phi_n \rangle$ is what,
for instance in Ref.\cite{myo98}, is called the response function, from which the strength function 
can be extracted according to Eq.(\ref{str}).

% BibTeX users please use
%\bibliographystyle{spbasic}

\begin{thebibliography}{3}
%\bibitem[Author I(1999)]{Ref1}
%Author I (year) Article title. Journal Title-Abbreviated Vol: pp--pp
%\bibitem[Author and Smith(2001)]{Ref2}
%Author I, Smith J (year) Book title. Publisher, Place, pp numbers
%% Format for proceedings
%\bibitem[Author and Smith(2003)]{Ref3}
%Author I, Smith J (year) Paper title. In: Editor, A. (ed.) Proceedings
%Title, Location, Date, pages. Publisher, Place
\bibitem{ald56} K. Alder, A. Bohr, T. Huus, B. Mottelson, and A. Winther, Rev. Mod. Phys. {\bf 28} (1956) 432.
\bibitem{lan86} K. Langanke and C. Rolfs, Phys. Rev. C {\bf 33} (1986) 790.
\bibitem{ber99} C.A. Bertulani, D.T. de Paula, and V.G. Zelevinsky, Phys. Rev. C {\bf 60} (1999) 031602.
\bibitem{myo98} T. Myo, A. Ohnioshi, and K. Kato, Prog. Theor. Phys. {\bf 99} (1998) 801.
\bibitem{ho83}  Y.K. Ho, Phys. Rep. {\bf 99} (1983) 1.

\bibitem{gar13} E. Garrido, A.S. Jensen, and D.V. Fedorov, 
Phys. Rev. C {\bf 86} (2013) 064608.          

\bibitem{gar13a} E. Garrido, A.S. Jensen, and D.V. Fedorov, submitted for publication.

\bibitem{tan85} O. Tanimura and U. Mosel, Nucl. Phys. A {\bf 440} (1985) 173.
\bibitem{zel60} Ya.B. Zel'dovich, Zh. Exp. Theor. Fiz. {\bf 39} (1960) 776.
\bibitem{lan86b} K. Langanke, Phys. Lett. B {\bf 174} (1986) 27.
\bibitem{buc77} B. Buck, H. Friedrich, and C. Wheatley, Nucl. Phys. A {\bf 275} (1977) 246.
\bibitem{ali66} S. Ali and A.R. Bodmer, Nucl. Phys. A {\bf 80} (1966) 99.
\bibitem{kro87} D. Krolle, H.J. Assenbaum, C. Funk, and K. Langanke, Phys. Rev. C {\bf 35} (1987) 1631.
\bibitem{til04} D.R. Tilley, J.H. Kelley, J.L. Godwin, D.J. Millener, J. Purcell, C.G. Sheu, and H.R. Weller,
                Nucl. Phys. A {\bf 745} (2004) 155.
\bibitem{lan86c} K. Langanke and C. Rolfs, Z. Phys. A {\bf 324} (1986) 307.
\bibitem{gre01} W. Greiner, {\it Quantum Mechanics: Special chapters}, Springer-Verlag (2001) pp 121.
\bibitem{moh94} P. Mohr, H. Abele, V. K\"{o}lle, G. Staudt, H. Oberhummer, and H. Krauss,
                Z. Phys. A {\bf 349} (1994) 339.
\bibitem{ber96} C.A. Bertulani, Z. Phys. A {\bf 356} (1996) 293.
\bibitem{nun99} F.M. Nunes and I.J. Thompson, Phys. Rev. C {\bf 59} (1999) 2652.
\bibitem{for03} C. Forss\'{e}n, N.B. Shul’gina, and M.V. Zhukov, Phys. Rev. C {\bf 67} (2003) 045801.
\bibitem{alo92} O.E. Alon and N. Moiseyev, Phys. Rev. A {\bf 46} (1992) 3807.
\bibitem{moi98} N. Moiseyev, Phys. Rep. {\bf 302} (1998) 247.
\bibitem{kat06} K. Kato, J. of Phys.: Conference Series {\bf 49} (2006) 73.
\bibitem{eco06} E.N. Economou, {\it Green's functions in quantum physics}, Springer Series in 
      Solid-State Sciences 7, Springer-Verlag, Berlin-Heidelberg-New York (2006) Chapter 1.
\bibitem{gir03} B. Giraud and K. Kato, Ann. of Phys. {\bf 308} (2003) 115.

\end{thebibliography}


\end{document}